\documentclass[12pt,preprint]{aastex}

\usepackage{lscape}

\newcommand{\dm}{\Delta{\rm m}_{15}(B)}
\shorttitle{Type Ia supernova distances to NGC~1316}
\shortauthors{Stritzinger et al.}
\usepackage{graphicx}

\received{2010 May XXX}
\accepted{2010  XXX XXX}
\begin{document}
\title{The Distance to NGC~1316 (Fornax~A) From Observations of 
Four Type~Ia Supernovae\altaffilmark{1}}

%`
\author{
Maximilian Stritzinger\altaffilmark{2,3,4},
Christopher R.~Burns\altaffilmark{5},
Mark~M.~Phillips\altaffilmark{2},
Gast\'on Folatelli\altaffilmark{6},
Kevin Krisciunas\altaffilmark{7,3},
ShiAnne Kattner\altaffilmark{8},
Sven~E.~Persson\altaffilmark{5},
Luis Boldt\altaffilmark{2},
Abdo Campillay\altaffilmark{2},
Carlos Contreras\altaffilmark{9},
Wojtek Krzeminski\altaffilmark{2},
Nidia Morrell\altaffilmark{2},
Francisco Salgado\altaffilmark{2},
Wendy~L.~Freedman\altaffilmark{5},
Mario Hamuy\altaffilmark{6},
Barry~F.~Madore\altaffilmark{5},
Miguel Roth\altaffilmark{2},
and
Nicholas~B.~Suntzeff\altaffilmark{7}
}

\altaffiltext{1}{
 This paper includes data gathered with the 6.5 meter Magellan telescope 
at Las Campanas Observatory, Chile.}

\altaffiltext{2}{Carnegie Observatories, Las Campanas Observatory, 
  Casilla 601, La Serena, Chile; {mstritzinger@lco.cl}.}

\altaffiltext{3}{Dark Cosmology Centre, Niels Bohr Institute, University 
of Copenhagen, Juliane Maries Vej 30, 2100 Copenhagen \O, Denmark; {max@dark-cosmology.dk}.} 

\altaffiltext{4}{
The Oskar Klein Centre, Department of Astronomy, Stockholm University, AlbaNova, 10691 Stockholm, Sweden; {max.stritzinger@astro.su.se}.}

\altaffiltext{5}{Observatories of the Carnegie Institution for
 Science, 813 Santa Barbara St., Pasadena, CA 91101, U.S.A.}

\altaffiltext{6}{Universidad de Chile, Departamento de Astronom\'{\i}a,
  Casilla 36-D, Santiago, Chile.}

\altaffiltext{7}{
George P. and Cynthia Woods Mitchell Institute for
Fundamental Physics and Astronomy, Texas A\&M University,
Department of Physics and Astronomy,  College Station, TX 77843, U.S.A.}

\altaffiltext{8}{Department of Astronomy, San Diego State University, San Diego, CA 92182-1221, U.S.A.}

\altaffiltext{9}{Centre for Astrophysics \& Supercomputing, Swinburne University of Technology, P.O. Box 218, Victoria 3122, Australia.}

\begin{abstract}
  \noindent 
The giant elliptical galaxy NGC~1316 (Fornax~A) is a well-studied member of the Fornax Cluster and a prolific producer of Type~Ia supernovae, having hosted four observed events since 1980.
Here we present detailed optical and near-infrared light curves of the spectroscopically 
normal SN~2006dd.  These data are used, along with previously published 
photometry of the normal  SN~1980N and SN~1981D, and the fast-declining,
low-luminosity SN~2006mr, to compute independent estimates of the host 
reddening for each supernova, and the distance  to  NGC~1316. 
From the three {\em normal} supernovae, we find a distance of 
17.8~$\pm$~0.3 (random) $\pm$~0.3 (systematic) Mpc for $H_o = 72$.
Distance moduli derived from the ``EBV" and Tripp methods
give values that are mutually consistent to 4 -- 8\%. % 2-3\%.
Moreover, the weighted 
means of the distance moduli for these three SNe for three methods agree to within 3\%.
This consistency is encouraging and supports the premise that Type~Ia
supernovae are reliable distance indicators at the 5\% precision level or better.
On the other hand, the two methods used to estimate the distance of the 
fast-declining SN~2006mr both yield a distance to NGC~1316 which is
25-30\% larger.
This disparity casts doubt on the suitability of fast-declining events for estimating extragalactic distances.
Modest-to-negligible host galaxy reddening values are derived for all four supernovae.
Nevertheless, two of them (SN~2006dd and SN 2006mr)
show strong Na~I~D interstellar lines in the host galaxy system.
The strength of this absorption is completely inconsistent with the small
reddening values derived from the supernova light curves if the gas in NGC~1316
is typical of that found in the interstellar medium of the Milky Way.
 In addition, the equivalent width of the Na lines in SN~2006dd appear to have weakened significantly
some 100-150~days after explosion.
 
\end{abstract}
   

\section{INTRODUCTION}
\label{sec:intro}

Type Ia supernovae  (hereafter SNe~Ia) are a powerful tool for measuring
cosmological parameters and characterizing the nature of dark energy; for a 
contemporary review see \citet{leibundgut08} and references therein.  Given 
their importance, it is sensible when the opportunity presents itself, to 
test the consistency of distances afforded by SNe~Ia.  NGC~1316 (Fornax~A) ---
a prolific producer of SNe~Ia --- offers such an opportunity, having hosted 
four Type~Ia events since 1980.  Here we present optical and near-infrared
(NIR) light curves of the normal Type~Ia SN~2006dd that cover the flux 
evolution from $-$10.5 to $+$228 days past $B$~band maximum
[T$(B$)$_{\rm max}$].  These data are used, along with previously published 
optical and NIR photometry of SNe~1980N, 1981D and 2006mr, to determine 
independent measures of the distance to NGC~1316.  Our analysis shows that 
the three {\em normal} SNe~Ia do deliver consistent distances to 
NGC~1316.  The fourth SN~Ia in NGC~1316, SN~2006mr, was a fast-declining,
sub-luminous event similar to the prototypical SN~1991bg 
\citep{filippenko92,leibundgut93,turatto96}.  Two methods are used to
estimate the distance of NGC~1316 using this SN, but the overall agreement
with the distance derived from the three normal SNe is poor.

NGC~1316 is a well-studied, radio-bright Morgan-D type galaxy in a peripheral 
area of the Fornax~I cluster. Detailed studies suggest that this galaxy is a 
remnant of an intermediate-age ($\sim$3 Gyr) merger 
\citep{schweizer80,goudfrooij01}.  The inner region is marked with a large 
dust lane while the outer region displays features related to tidal perturbations. 
Two giant radio lobes powered by a supermassive black hole extend from the 
center \citep{schweizer80,geldzahler84,nowak08}.

According to the infrared dust maps of \citet{schlegel98}, the 
reddening due to dust within the Milky Way in the direction of NGC~1316 
is $E(B-V)_{\rm gal} = 0.021$~mag.  As we show below, the color 
excesses derived from the four SN~Ia studied here, two of which lie in 
close proximity to the dust in the inner region of NGC~1316, suggest that 
they suffered relatively modest-to-negligible amounts of reddening.
  
The organization of this paper is as follows:  $\S$~2 describes 
the observational data set, $\S$~3 presents the analysis where we 
estimate the host reddening and derive distances to NGC~1316 for each 
SN, $\S$~4 discusses the strong interstellar Na~I~D absorption lines observed 
in the spectra of SNe~2006dd and 2006mr, and $\S$~5 summarizes the 
conclusions. 

\section{The Observational Sample}

\begin{figure}
\figurenum{1}
\epsscale{0.8}
\plotone{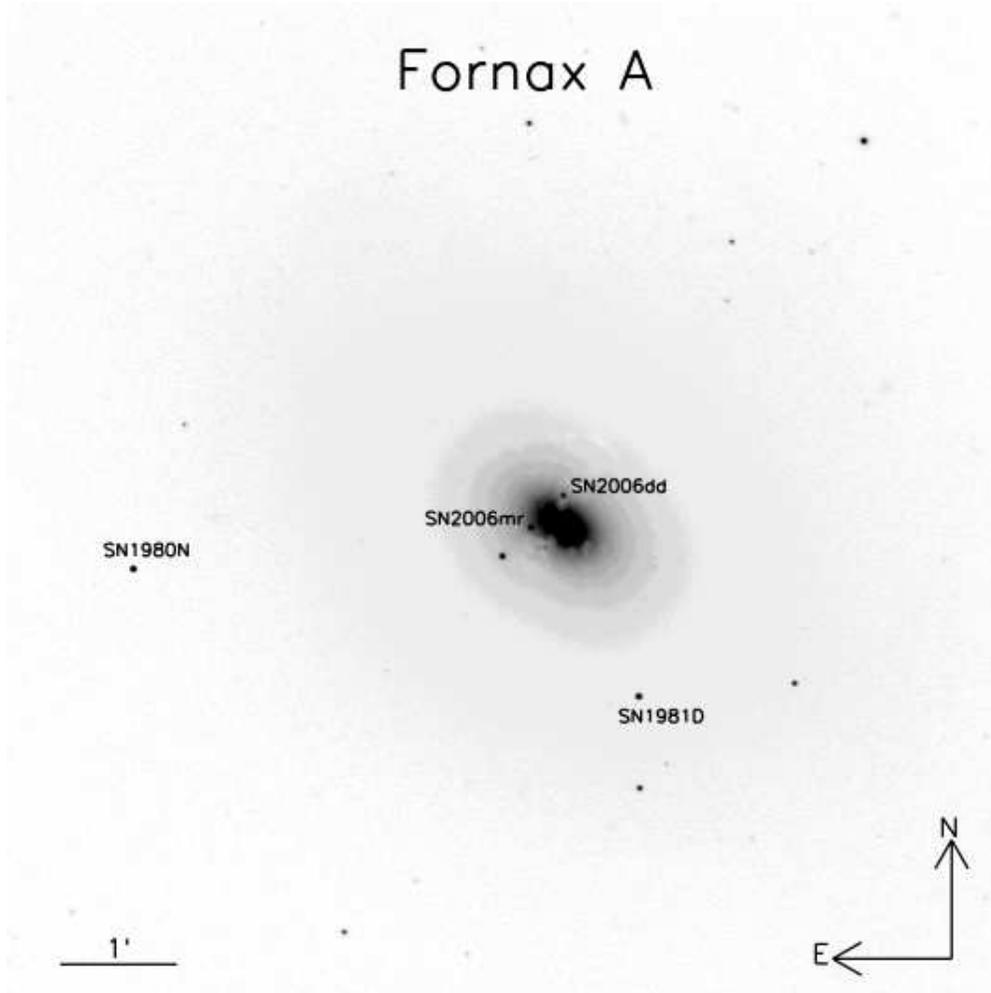}
\caption[Fig1.eps]{A Swope 1~m telescope $V$-band image of NGC~1316 obtained in 
2006.  Artificial stars have been placed at the coordinates of SN~1980N and 
SN~1981D.\label{fig:FC}}
\end{figure}

The four SNe~Ia that are the subject of this paper are SN~1980N, SN~1981D, 
SN~2006dd, and SN~2006mr.  Figure~\ref{fig:FC}  displays an image of 
NGC~1316 with the location of each SN indicated.  SN~1980N was discovered 
on the rise by \citet{maza80} and later identified as a Type~Ia event 
\citep{blanco80,prabhu81}.  Barely three months passed before a second SN,
1981D, was discovered in NGC~1316 \citep{cragg81,evans82} and was also 
classified as a SN~Ia \citep{menzies81}.  Optical spectra of SN~1980N 
published by  \citet{prabhu81} and \citet{hamuy91} reveal that this was a 
spectroscopically normal SN~Ia.  To our knowledge, a spectrum of SN~1981D has
never been published, but the description of \citet{menzies81} is consistent
with it also having been a normal SN~Ia.

Nearly a quarter of a century later, on 2006 June 19.2 UT, \citet{monard06a} 
reported the appearance of SN~2006dd in the central region of the galaxy,
that was spectroscopically classified as a young SN~Ia  
\citep{salvo06,morrell06}.  Less than five months later, \citet{monard06b} 
discovered yet another SN in the central region of 
NGC~1316.  Soon afterwards, this event, designated SN~2006mr, was 
reported to be a SN~Ia similar to the fast-declining 
SN~1986G \citep{phillips06}.

\begin{figure}
\figurenum{2}
\epsscale{1.0}
\plotone{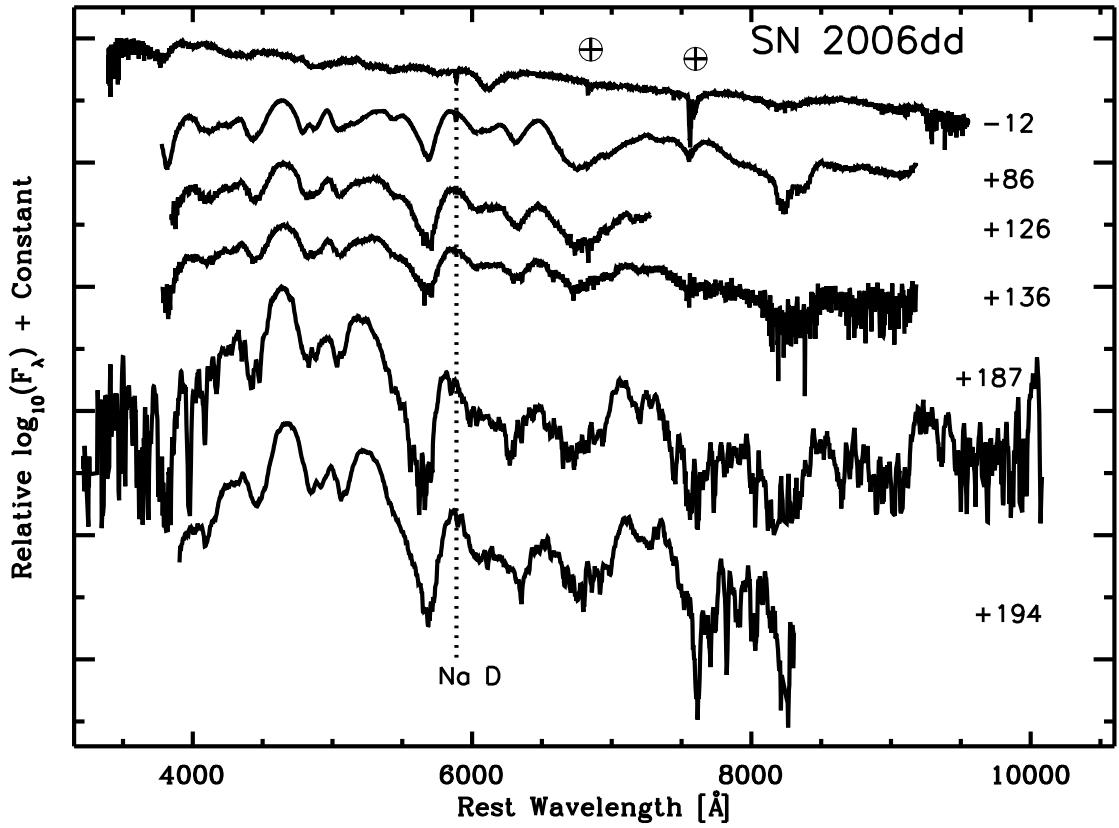}
\caption[Fig2.eps]{Early and late phase optical spectroscopy of SN~2006dd. Telluric
features are indicated with an Earth symbol.}\label{fig:SN06ddspectra}
%%{\center Stritzinger {\it et al.} Fig. \ref{fig:FC}}
\end{figure}

\begin{figure}
\figurenum{3}
\epsscale{1.0}
\plotone{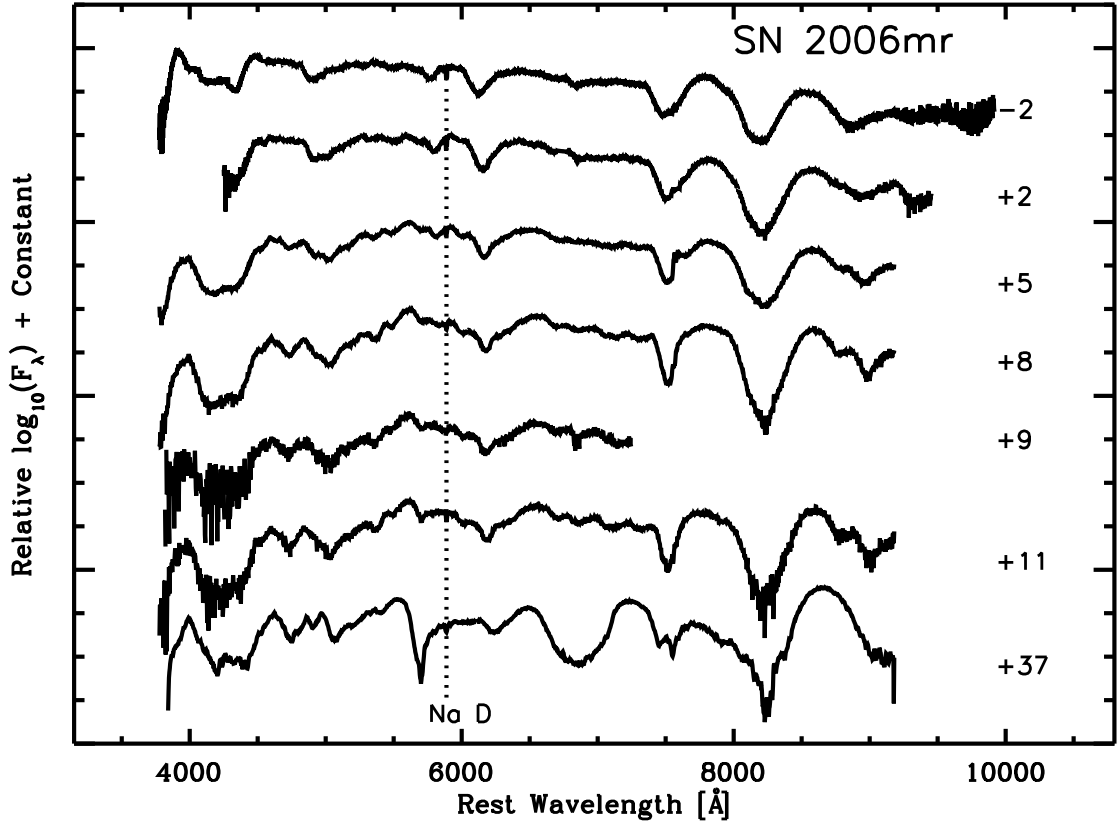}
\caption[Fig3.eps]{Optical spectroscopy of SN~2006mr.\label{fig:SN06mrspectra}}
%%{\center Stritzinger {\it et al.} Fig. \ref{fig:FC}}
\end{figure}

\begin{figure}
\figurenum{4}
\epsscale{1.0}
\plotone{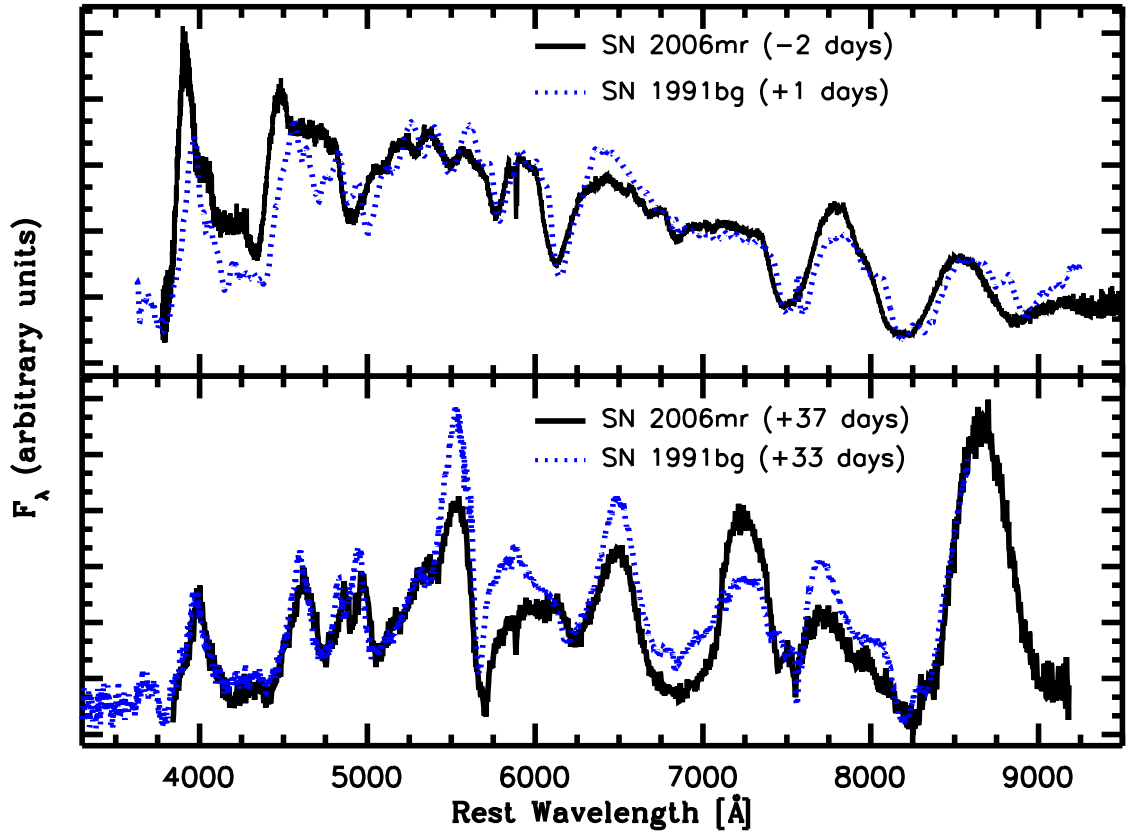}
\caption[Fig4.eps]{Comparison of optical spectra of SN~1991bg and SN~2006mr.
The spectra of SN~1991bg are from \citet{turatto96} and were obtained from the 
SUSPECT database.\label{fig:spectracompare}}
%%{\center Stritzinger {\it et al.} Fig. \ref{fig:FC}}
\end{figure}

Optical spectroscopy of SNe 2006dd and 2006mr was obtained during the course 
of the Carnegie Supernova Project \citep[CSP; ][]{hamuy06}.  
Table~\ref{tab1} contains a journal of the spectroscopic observations, and the spectra 
themselves are shown in Figures~\ref{fig:SN06ddspectra} and \ref{fig:SN06mrspectra}.  
The six spectra of SN~2006dd displayed in  
Figure~\ref{fig:SN06ddspectra} range from $-$12 to 194 days past 
T($B$)$_{\rm max}$, and both the early- and late-phase spectra resemble that of 
a normal SN~Ia.  The spectroscopic sequence of SN~2006mr covers from $-$2 
to $+$63 days past T($B$)$_{\rm max}$, and 
displays a striking resemblance to 
the fast-declining, sub-luminous SN~1991bg \citep{filippenko92,leibundgut93,turatto96}. 
This is demonstrated in the comparison of near- and post-maximum spectra shown in 
Figure~\ref{fig:spectracompare}.
The spectra of both SN~2006dd and SN~2006mr exhibit conspicuous Na~I~D 
interstellar absorption at the redshift of NGC~1316.
As discussed in $\S$~4, there is evidence that the equivalent width of the 
Na~I~D absorption in SN~2006dd varied during the course of our 
spectroscopic coverage.

In the remainder of this section we discuss the optical and NIR broad-band 
observations of these four SN~Ia.
  
\subsection{Supernova 1980N} 

Nearly three weeks worth of optical photoelectric photometry was obtained by
Landolt \citep{hamuy91} and \citet{olszewski82}.  A detailed analysis of 
these data is presented in \citet{hamuy91}.  The $BVRI$ light curves cover 
the flux evolution from $-$0.8 to $+$59.1 days past T$(B)_{\rm max}$, and 
form a consistent set of photometry on the Landolt system. 

To  check the assumption made in \citet{hamuy91} that the host galaxy 
background had a negligible impact on the photometry, we measured the 
brightness of the host at the position of SN~1980N in high signal-to-noise 
images constructed from a series of frames obtained by the CSP at Las Campanas
Observatory (LCO) throughout 2006.  Prior to reassessing the background, we 
first determined new coordinates of SN~1980N. This was done in the following 
manner.  Figure~1 of \citeauthor{hamuy91} was scanned and converted 
to a {\tt FITS} image from which the {\tt XY} coordinates of a number of 
field stars were used to compute an astrometric world coordinate solution 
(WCS).  Furnished with the WCS and the {\tt XY} positions of  SN~1980N, we 
then generated new 2000.0 coordinates of 
$\alpha =$03$^{\rm h}$23$^{\rm m}$00${\fs}$32 and 
$\delta$ $= -$37$\arcdeg$12$\arcmin$49$\farcs$9.  We conservatively estimate
the error in this position to be 2--3$\arcsec$ since an extrapolation of
the astrometric solution was required.  Note, however, that these new 
coordinates differ by 10$\arcsec$ in $\alpha$ and 2$\arcsec$ in $\delta$
from the position of SN~1980N given in the NASA Extragalactic Database (NED),
which are from \citet{tsvetkov93}.

Next, at this location, a 31$\arcsec$ aperture was used to measure 
$B$- and $V$-band magnitudes from the 2006 stacked images of 18.718~$\pm$~0.054  mag
and 17.797~$\pm$~0.032  mag, respectively. These values are $\sim$3 magnitudes 
fainter than the faintest SN measurement reported in \citet{hamuy91}.
Moreover, Landolt probably used a 16$\arcsec$ aperature or smaller for his
observations.  This means that the galaxy light in his aperture was at least
60 times fainter than the SN, confirming that no background correction is
necessary for the optical photometry of SN~1980N. 

\citet{elias81} presented $JHK$ light curves of SN~1980N that follow 
the flux evolution from day  $+$5.3 to $+$72.2.  These measurements were
made with single channel photometers attached to telescopes at Cerro Tololo 
Inter-American Observatory (CTIO)\footnote[10]{Cerro Tololo Inter-American 
Observatory, Kitt Peak National Observatory, National Optical Astronomy 
Observatories, operated by the Association of Universities for Research in 
Astronomy, Inc., (AURA), under cooperative agreement with the National Science 
Foundation.} and LCO.  The host galaxy contamination was estimated by sky
observations made near the position of the SN.  The final magnitudes were
published on the CIT/CTIO system \citep{elias82}.

Plotted in Figure~\ref{fig:SN80NLCS} are the optical and NIR light curves of 
SN~1980N.
  
\subsection{Supernova 1981D}

\citet{walkermarino82} published nine epochs of $B$- and $V$-band 
photoelectric photometry that ranged from day $-$4.6 to $+$23.3.  
These observations were obtained on an unknown photometric 
system through a 31$\arcsec$ aperture, with no correction for the host 
galaxy light.  As described above for SN~1980N, we measured the 
brightness of the host galaxy at the position of SN~1981D using the high 
signal-to-noise images taken in 2006.  First, the 2000.0 coordinates of 
SN~1981D were determined to be $\alpha =$03$^{\rm h}$22$^{\rm m}$38${\fs}$38 
and $\delta$ $= -$37$\arcdeg$13$\arcmin$57$\farcs$8 using the same 
method described for SN~1980N.  The error in this 
position is $\sim$1~$\arcsec$.  Again, these new coordinates are 
substantially different (15$\arcsec$ in $\alpha$ and 9$\arcsec$ in $\delta$)
than those listed in NED \citep{tsvetkov93}.
The galaxy background at this position was measured in the  2006 stacked images with
a 31$\arcsec$ aperture  to be $B = 15.230\pm0.010$ mag  and $V = 14.280\pm0.030$  mag. 
These robust measurements are brighter by 0.196~mag 
in $B$ and 0.218~mag in $V$ than those used in \citet{hamuy91} to correct 
the \citet{walkermarino82} photometry for background contamination.  We 
therefore returned to the original \citeauthor{walkermarino82} photometry 
and corrected for  the new estimates of the background at the location of SN~1981D.

Seven epochs of NIR photometry of SN~1981D obtained at CTIO and LCO were
also published by \citet{elias81}. These $JHK$ light curves follow the 
evolution from day $+$4.2 to $+$16.0, and are corrected 
for the host galaxy background light.  The published magnitudes are in
the CIT/CTIO system.

The optical and NIR light curves of SN 1981D are
plotted in Figure~\ref{fig:SN81DLCS}. 

\subsection{Supernova 2006dd}

Thirteen epochs of optical and NIR imaging of SN~2006dd were obtained at 
CTIO with the SMARTS consortium 1.3~m telescope equipped with ANDICAM.  The 
resulting $BVRIJHK_s$ light curves nicely sample the flux evolution from 
$-$10.5 to $+$55.4 days past T$(B)_{\rm max}$.  Additional optical ($uBgVri$) 
imaging was carried out by the CSP with the 
Henrietta Swope 1.0~m telescope and SITe3 CCD camera at LCO. These light 
curves span from day $+$60.5 to $+$228.2.  A small amount of NIR imaging
was also obtained by the CSP with the RetroCam imager on the Swope telescope
after day $+$61.  Complete details regarding the observing procedures used 
with ANDICAM, SITe3, and RetroCam can be found in \citet{krisciunas09}, 
\citet{hamuy06}, and \citet{contreras10}.

All optical images were reduced in a standard manner including: 
(1) bias subtraction, (2) flat-field  division, and (3) the application of 
a shutter time correction.  NIR images were also reduced in a standard manner, 
consisting of (1) dark subtraction, (2) flat-field division,
(3) sky subtraction, and (4) geometric alignment and combination of
the dithered frames.  Further details of the data reduction procedures
are given in \citet{krisciunas09}, \citet{hamuy06}, and \citet{contreras10}.

Photometry of SN~2006dd was computed differentially with respect to a 
sequence of stars in the field of NGC~1316.  The local sequence used to 
compute optical photometry from the ANDICAM images consists of four stars 
calibrated to observations of the PG1657 field \citep{landolt92} obtained on 
five photometric nights.  Final magnitudes for these stars are given in
Table~\ref{tab2}.  For the SITe3 $uBgVri$ imaging, we adopted the 
same local sequence established in \citet[][see their Table~3]{contreras10} 
to compute photometry of SN~2006mr (see below).  This sequence was calibrated 
to the \citet{landolt92} $BV$ and \citet{smith02} $ugri$ systems through 
standard star observations obtained over the course of four photometric nights. 

Three stars are in common between the two different local 
sequences.  Two of these stars have consistent $B$ and $V$ magnitudes,
however the faintest of these three stars exhibits a difference of up to nearly
a tenth of a magnitude. The average difference between the three common stars
is found to be 0.038 mag and 0.042 mag in $B$ and $V$, respectively. 

Prior to computing final photometry, template images of NGC~1316 were 
subtracted at the location of SN~2006dd from all science images.  The 
templates were obtained well after the SN faded, with the identical 
telescope/instrument/filter combinations used to obtain the science images.
Finally, point spread function (PSF) photometry was computed from the 
template-subtracted science images.  

To facilitate the analysis of the light curves (see $\S$~3), we elected 
to forego the application of color corrections, and computed the ANDICAM 
optical photometry in the {\em natural} system of the 1.3~m telescope.  To 
accomplish this, it was necessary to convert the optical magnitudes listed 
in Table~\ref{tab2} from the standard to the natural system using the
following equations:

\begin{equation}
B_{\rm nat} = B_{\rm std} - CT_{B} \cdot(B_{\rm std} - V_{\rm std})
\label{beqn}
\end{equation}
\begin{equation}
V_{\rm nat} = V_{\rm std} - CT_{V} \cdot(B_{\rm std} - V_{\rm std})
\label{veqn}
\end{equation}
\begin{equation}
R_{\rm nat} = R_{\rm std} - CT_{R} \cdot(V_{\rm std} - R_{\rm std})
\label{reqn}
\end{equation}
\begin{equation}
I_{\rm nat} = I_{\rm std} - CT_{I} \cdot(V_{\rm std} - I_{\rm std}).
\label{ieqn}
\end{equation}

\noindent These transformations adopt the mean color terms derived from 
ANDICAM observations of standard fields obtained during seven photometric 
nights between June~22 to August~25, 2006.  Specifically, the color terms 
used are CT$_{B}$ $=$ $+$0.091$\pm$0.014, CT$_{V}$ $=$ $-$0.019$\pm$0.007, 
CT$_{R}$ $=$ $+$0.031$\pm$0.011, and CT$_{I}$ $=$ $-$0.032$\pm$0.012. 
The resulting ANDICAM photometry of SN~2006dd on the natural system of the 
1.3-m telescope is given in Table~\ref{tab3}.

Likewise, the LCO optical photometry of SN~2006dd was computed in the natural 
system of the Swope 1~m telescope $+$ SITe3 camera following the procedures 
detailed in \citet{contreras10}.  Final magnitudes are listed in 
Table~\ref{tab4}.  

The NIR photometry of SN~2006dd obtained with ANDICAM and RetroCam was reduced 
in a similar fashion to the optical data.  The $JHK_s$ magnitudes of star~CSP02 of
the ANDICAM local sequence were calibrated directly to the \citet{persson98} 
NIR photometric system via observations of the standard stars P9109, P9150, 
and P9172 assuming no color term.  This star is also 
in the 2MASS catalog, and we find that our magnitudes are between 0.02 to 
0.06~mag brighter than their 2MASS values.  Table~\ref{tab2} contains the 
final $JHK_s$ magnitudes of star~CSP02 in the Persson standard system.  For the 
CSP RetroCam observatons on the Swope 1~m telescope, we adopted the local NIR 
sequence for SN~2006mr published by \citet[][see their Table~3]{contreras10}.
This sequence was calibrated to the \citet{persson98} system through
standard star observations obtained over the course of 15 photometric nights.
Final NIR magnitudes for SN~2006dd in the natural systems of the 1.3~m and
Swope 1~m telescope are given in Table~\ref{tab3}.

Plotted in Figure~\ref{fig:SN06ddLCS} are the early phase optical and NIR 
light curves of SN~2006dd computed from the ANDICAM and RetroCam images.  In
Figure~\ref{fig:SN06ddLCS2}, the full set of $BVri$ observations obtained
at CTIO and LCO are shown together.  ``S-corrections'' \citep{suntzeff00,stritzinger02}
calculated using an updated version of the \citet{hsiao07} SN~Ia spectral
template have been applied to the ANDICAM observations to place them on the 
natural system of the Swope telescope.  Although there is no overlap between 
the two sets of observations, the agreement is generally quite good. 

The later phase light curves of SN~2006dd are amongst the most densely 
sampled yet obtained for a SN~Ia out to day $+$230. Linear least-squares 
fits to the photometry after 150~days past T($B$)$_{\rm max}$ yield decline 
rates of 1.6, 1.6, 2.2, and 1.5~mag per hundred days in $B$, $V$, $r$, and $i$.
These decline rates are marginally steeper than what has been 
documented in most other SNe~Ia \citep{lair06}.

\subsection{Supernova 2006mr}

Detailed optical ($uBgVri$) and NIR ($YJH$) light curves of SN~2006mr
have recently been published by the CSP \citep{contreras10}. The photometry 
spans from $-$4.3 to $+$96.6 days past T($B)_{\rm max}$, and is on the
Swope telescope natural system.  SN~2006mr is one of several low-luminosity,
fast-declining events contained in the first CSP data release of SNe~Ia, 
and a detailed analysis of these data is presented in \citet{folatelli10}.

\section{Analysis}

\subsection{Light Curve Fitting}

To estimate the reddening and distance to each SN~Ia, we  made use of the 
template light curve fitting package SNooPy [SuperNovae in object-oriented 
Python; \citet{burns10}].  A description of SNooPy and the construction of 
its template light curves is presented in \citet{folatelli10}; here we 
briefly summarize. 

SNooPy is based on a modified version of the Prieto method \citep{prieto06},
and contains optical ($uBgVri$) and NIR ($YJH$) template light 
curves built from well-observed SNe~Ia obtained during the first four 
campaigns of the CSP.  To fabricate the templates, cubic spline functions 
are fitted to each light curve (corrected for time dilation) and estimates of 
the peak magnitude, $m_{\rm max}$, the time of maximum, T$_{\rm max}$, and 
the decline rate parameter, $\dm$, are extracted.\footnote[11]{$\dm$ is defined 
as the difference between the $B$-band magnitude at peak and 15 days later.  
This parameter correlates with the absolute peak magnitude of a SNe~Ia in the 
sense that more luminous events have smaller decline rates \citep{phillips93}.
Because a single value of $\dm$ is used in SNooPy to parameterize all the 
filters of a SN~Ia, the values reported here will differ somewhat from the value 
one would get by measuring it directly from the decline of the $B$-band 
light-curve alone.} The values of these parameters are then combined with the 
\citet{hsiao07,hsiao10} SN spectral templates, and K-corrections are 
computed \citep[see][$\S$~5.1 for more details]{freedman09}.  The 
K-corrected photometry then undergoes another round of spline fits yielding
revised estimates of $m_{\rm max}$, T$_{\rm max}$, $\dm$, as well as  
a complete covariance matrix for the light curve parameters. 
 Each filter's K-corrected and time-dilation-corrected photometric data now form a 2D surface in a 3D
  [t--T$_{\rm max}$, m-m$_{\rm max}$, $\dm$] space, which
can generate for any given $\dm$ a template light 
curve by simply interpolating within this surface.  
 This manner of interpolation is the main difference from the original Prieto method, and boils down to SNooPy producing templates by doing 2D interpolating among the original data points of the SNe rather than using the weighted average of the 1D spline functions.

The scarcity of published $K_s$-band light curves with decent pre-maximum 
sampling hinders our ability to build suitable templates for this particular 
passband. We therefore fit the observed photometry with the $K_s$-band 
polynomial template presented by \citet{krisciunas04}, multiplied by 
an appropriate stretch value that is related to $\dm$ 
\citep[see][Equation 1]{folatelli10}. 

As the template light curves built into SNooPy are on the natural system of 
the Swope telescope, it is necessary to transform (when appropriate) the SN 
photometry to be fit by SNooPy to this system.  This was accomplished through 
the application of  time-dependent corrections computed synthetically from 
template SN spectra \citep{suntzeff00}.  Done correctly, these 
``S-corrections" also require an accurate idealization of the total response 
function of the telescope/instrument/filter combination used to obtain the 
science images.  We therefore carefully constructed model passbands 
where necessary,  and then, following the method presented in 
\citet{stritzinger02}, combined these passbands with a library of template 
SN spectra \citep{hsiao07,hsiao10} to compute the S-corrections.  As mentioned in 
$\S$~2.2, it is not known in which photometric system the $B$- and $V$-band light 
curves of SN~1981D are published.  In what follows, we simply assume that 
they are on a photometric system similar to that of the Swope telescope, i.e.,
assume $B$ and $V$ are in the CSP natural system and require no S-corrections.  
If we were to assume these bands are on the standard Landolt system, the
estimated distance modulus to NGC~1316 decreases  by 
only $\sim$0.002 mag.

SNooPy has two different models for performing the light curve fits.
In the ``max'' model, the observed magnitudes of the SN in each filter are fitted
simultaneously according to the following equation:

\begin{eqnarray}
m_{X}[t-T(B)_{\rm max}] & = & T_X \big[(t-T(B)_{\rm max})/(1+z),\dm]\;+\;m_X^{\rm max}.
\label{beqn5}
\end{eqnarray}

\begin{figure}
\figurenum{5}
\epsscale{1.0}
\plotone{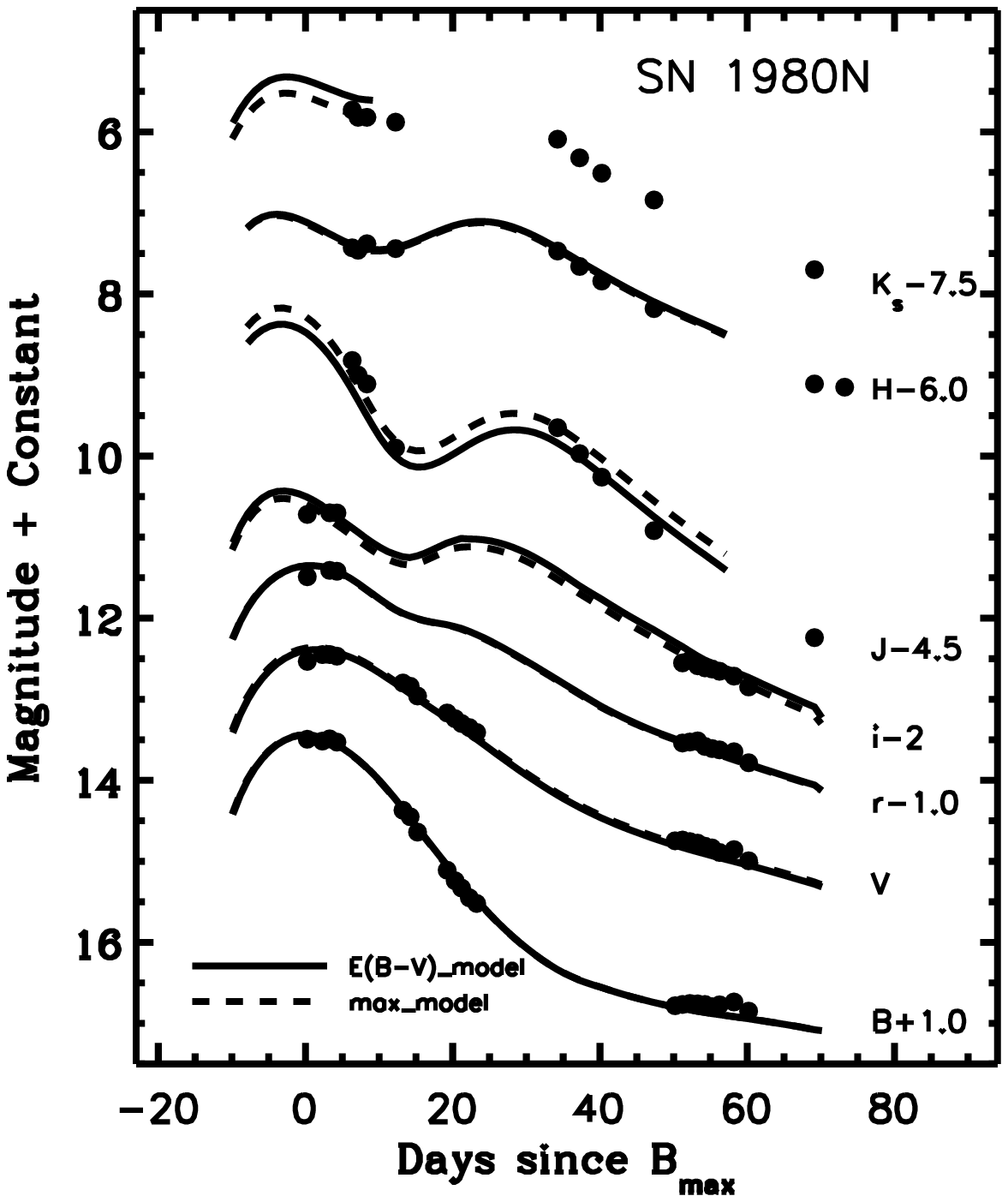}
\caption[Fig5.eps]
{$BVriJHK_s$ light curves of SN~1980N on the natural system
of the Swope 1~m telescope.
The smooth curves are the best SNooPy ``EBV model" fits (solid lines) and
the dashed lines correspond to the best ``max model" fits. 
Note that  the light 
curves have been shifted in the y-direction for clarity. 
\label{fig:SN80NLCS}}
%%{\center Stritzinger {\it et al.} Fig. \ref{fig:FC}}
\end{figure}

\begin{figure}
\figurenum{6}
\epsscale{1.0}
\plotone{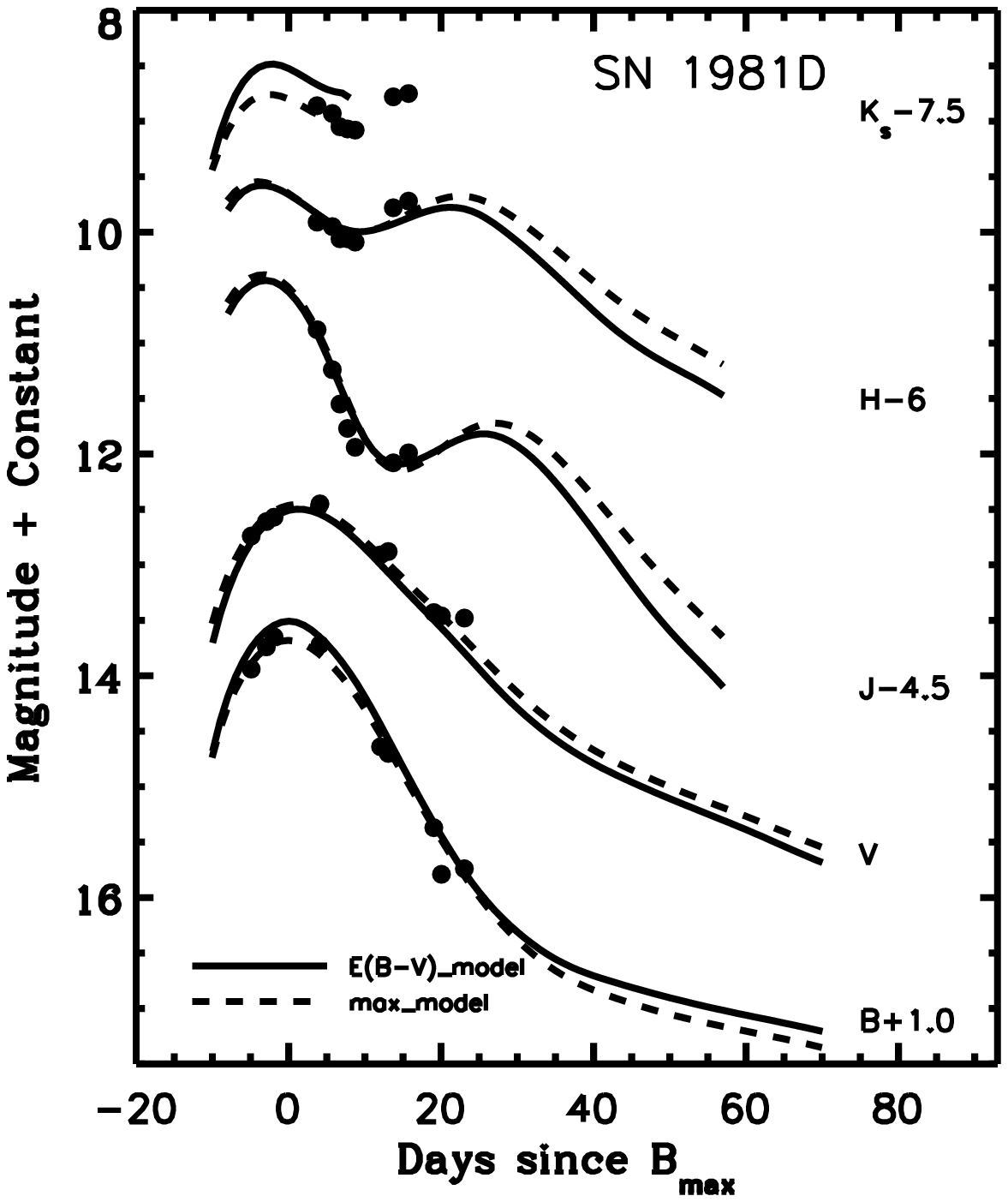}
\caption[Fig6.eps]
{$BVJHK_s$ light curves of SN~1981D on the natural system of the Swope 1~m
telescope. 
The smooth curves are the best SNooPy ``EBV model" fits (solid lines) and
the dashed lines correspond to the best ``max model" fits. 
Note that the light 
curves have been shifted in the y-direction for clarity. 
\label{fig:SN81DLCS}}
%%{\center Stritzinger {\it et al.} Fig. \ref{fig:FC}}
\end{figure}

\begin{figure}
\figurenum{7}
\epsscale{1.0}
\plotone{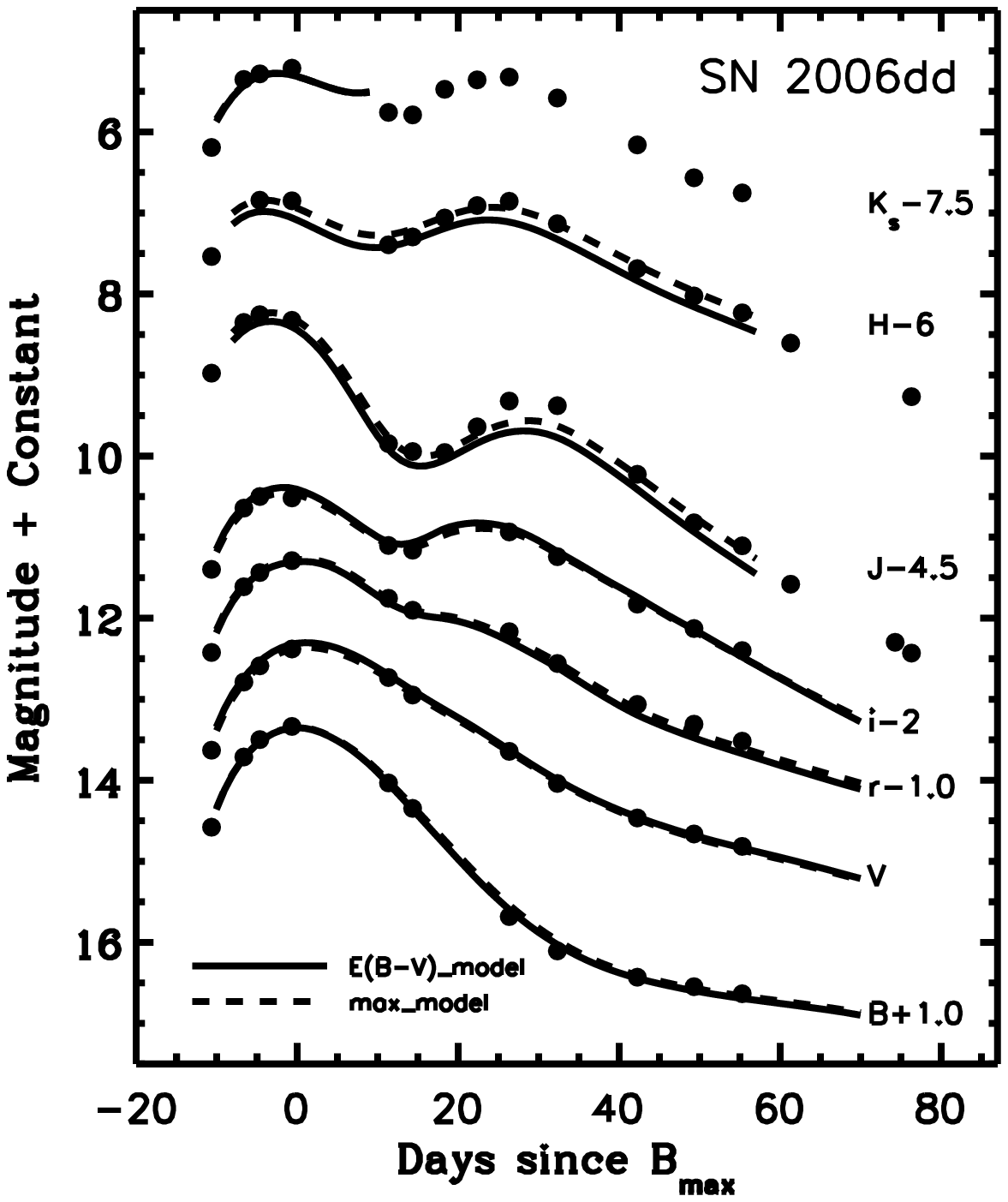}
\caption[Fig7.eps]
{ $BVriJHK_s$ light curves of SN~2006dd on the natural system of the Swope 1~m
telescope. 
The smooth curves are the best SNooPy ``EBV model" fits (solid lines) and
the dashed lines correspond to the best ``max model" fits. 
Note that  the light 
curves have been shifted in the y-direction for clarity. 
\label{fig:SN06ddLCS}}
%%{\center Stritzinger {\it et al.} Fig. \ref{fig:FC}}
\end{figure}

\begin{figure}
\figurenum{8}
\epsscale{1.0}
\plotone{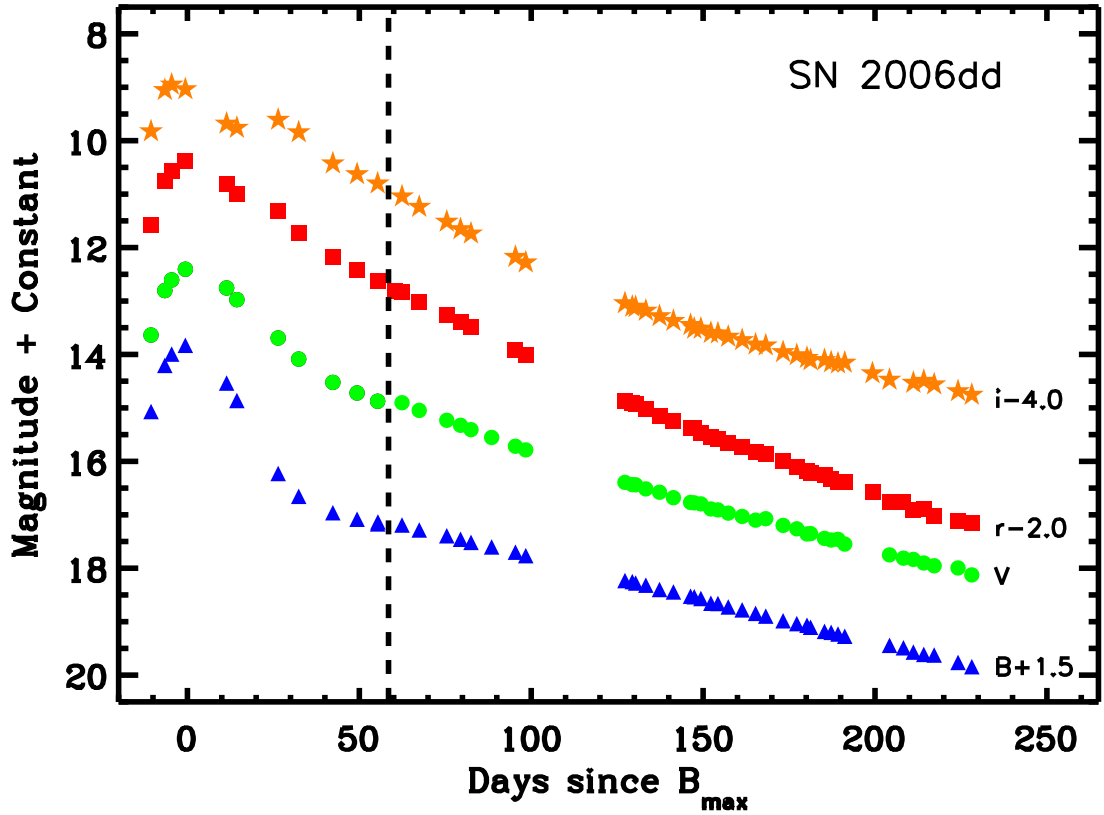}
\caption[Fig8.eps]
{Optical $BVri$ light curves of SN~2006dd on the natural system of the 
Swope 1~m telescope. The vertical line indicates the epoch between the
ANDICAM and Swope photometry. After the application of the S-corrections to the ANDICAM 
photometry to put it on the natural system of the Swope telescope, the photometry is in 
excellent agreement with the Swope photometry in the $Bri$ bands, but less 
consistent in the $V$ band.
\label{fig:SN06ddLCS2}}
%%{\center Stritzinger {\it et al.} Fig. \ref{fig:FC}}
\end{figure}

\noindent The input data for this model are the redshift of the host galaxy, $z$,
the observed magnitudes in filter $X$, $m_X$, as a function of time, $t$, 
corrected for Galactic reddening assuming a \citet{cardelli89} reddening law
with $R_V = 3.1$ and K-corrected as described in \citet{folatelli10}.
The fitting of the photometry to the SNooPy template functions, $T_X$, is performed 
simultaneously in all observed bands through $\chi^2$ minimization, yielding as 
final output parameters: $T(B)_{max}$, 
$\dm$, and the peak magnitudes, $m_X^{\rm max}$, in each filter.  The results of 
these fits for SNe~1980N, 1981D, and 2006dd are given in Table~\ref{tab5} and 
plotted as dashed lines on top of the observed photometry in 
Figures~\ref{fig:SN80NLCS}--\ref{fig:SN06ddLCS}.
The quoted uncertainties for the fit parameters were derived from the 
covariance matrix of the best template fits, scaled such that $\chi^{2}_{\nu} $$= 1$.

Unfortunately, SNooPy does not yet do a very good job of fitting the light curves 
of fast declining SNe~Ia.  Hence, for SN~2006mr, we take the light curve parameters 
given by \citet{folatelli10} that were estimated through the use of spline fits.  
These values are reproduced in Table~\ref{tab5}.  Note that, in this case,
the parameter, $\dm$, is a direct measurement of the decline rate of the
$B$ light curve.

Using the ``EBV'' model in SNooPy, the light curve data are 
fit to the following alternative model:

\begin{eqnarray}
m_{X}[t-T(B)_{\rm max}] & = & T_X\big[(t-T(B)_{\rm max})/(1+z),\dm]\;+\;M_X^{\rm max}[\dm]\nonumber\\
& & +\;\mu_\circ\;+\;R_X^{\rm Host} \cdot E(B-V)_{\rm Host}
\label{beqn6}
\end{eqnarray}

\noindent Here $M_X^{\rm max}$[$\dm$] is the unreddened 
absolute magnitude at maximum in band $X$ of a SN~Ia with a decline rate 
of $\dm$, 
$E(B-V)_{\rm host}$ is the reddening produced in the host 
galaxy, and $R_{X}^{\rm host}$ is the total-to-selective absorption 
coefficient.   
For the values of $M_X^{\rm max}$[$\dm$] and
$R_{X}^{\rm host}$, we adopt the recommended calibration from 
\citet{burns10}, which uses a subsample of unreddened SNe~Ia from
\citet{folatelli10} and excludes the two very red SNe~Ia SN~2006A and
SN~2006X.  
Note that this calibration assumes H$_{\circ} = 72$ km s$^{-1}$ Mpc$^{-1}$.
Simultaneously
fitting the light curve data in all filters yields final output values of $T(B)_{\rm max}$, 
$\dm$, $E(B-V)_{\rm host}$, and the 
distance modulus, $\mu_\circ$.   Best-fit models for SNe~1980N, 1981D, and 
2006dd derived from Equation~\ref{beqn6} are plotted as smooth curves 
 in Figures~\ref{fig:SN80NLCS}--\ref{fig:SN06ddLCS}, and the pertinent
fit parameters are given in Tables~\ref{tab6} and \ref{tab7}.  
The quoted values of 
$\mu$$_{0}$ listed in Table~\ref{tab7} were computed 
by solving for the single value of $\mu_\circ$ that minimizes the $\chi^2$ for all of the light curves.
As the \citet{burns10} relations are only valid for SNe~Ia with decline rates in the range
$0.7 < \dm < 1.7$, the EBV model cannot be applied to
the fast-declining SN~2006mr. 

A peculiarity of the \citet{burns10} calibration is its low value for the host-galaxy's total-to-selective absorption 
coefficient, $R_{V}^{\rm host} \sim 2$.  Indeed, most of the derived values of $R_{V}^{\rm host}$  from \citet{folatelli10} are very low
compared to the Milky Way.  It can be argued that if interstellar dust in these hosts is anything like 
that found in the Milky Way or in the Small and Large Magellanic clouds, such a value of $R_{V}^{\rm host}$  is physically unreasonable.  However, when considering only the ratio of the color excesses in different filters, 
\citet{folatelli10} derive an average value of $R_{V}^{\rm host}$  that is consistent with $R_{V}^{\rm host} =  3.1$.  Nevertheless, 
imposing a larger value of $R_{V}^{\rm host}$  results in poorer quantitative fits
to the light curves of those SNe~Ia in the calibration sample 
with large extinctions, as well as a much larger dispersion in the derived Hubble diagram.  
A simple explanation is that there is  another physical process, other than dust, causing intrinsically 
red SNe to be  dimmer and that this process is uncorrelated with $\dm$.  
The {\em empirical} fact remains 
that using a small value of $R_{V}^{\rm host}$ results in better standard candles,
especially for SNe~Ia that occur in massive host galaxies such as NGC~1316 \citep{sullivan10}.

\subsection{Host Galaxy Reddening}

The $E(B-V)$ color excesses derived from the SNooPy EBV method (listed in 
the second column of Table~\ref{tab6}) 
indicate that the three ``normal'' events, SNe~1980N, 1981D, and 2006dd, suffered 
modest to negligible amounts of host galaxy reddening. 
As previously mentioned, the calibration from \citet{burns10} is not
applicable to the fast-declining, intrinsically-red SN~2006mr, and so its 
host galaxy reddening cannot be estimated via the SNooPy fits. 
 However,
\citet{lira95} found that the $(B-V)$ colors of SNe~Ia at 30--90 days past $V$ 
maximum evolve in a nearly identical fashion, offering an alternative
method for measuring host galaxy reddening.  Lira's sample included
only one fast-declining event (SN~1991bg), but
 \citet[][see also \citet{garnavich04}]{taubenberger08} studied a 
sample of eight fast-declining events and concluded that they all follow 
reasonably well the Lira relation \citep{lira95}.  Applying the Lira relation 
to SN~2006mr, \citet{folatelli10} derived a color 
excess of  $E(B-V)_{\rm host} = -0.025~\pm~0.013$  mag.  This estimate 
indicates that SN~2006mr suffered essentially no host reddening.
For completeness, we have applied the \citet{folatelli10} calibration of the Lira law to
the two normal events with sufficient late-time coverage, SNe~1980N and 2006dd.
The results (given in the third column of Table~\ref{tab6}) are
fully consistent with the modest reddening values yielded by SNooPy.

An alternative way to estimate the host reddening is through
the  comparison of ``zero-reddening''  loci to the
observed $V$ minus NIR colors of each supernova.
\citet{krisciunas00} have provided such a set of low zero-reddening loci,
but these were fabricated in part with photometry  of  
SNe~1980N and 1981D.
 This motivated us to construct a new set of zero-reddening loci
 based on photometry of the well-observed SN~2001el \citep{krisciunas03}. 
 To obtain these, the observed  $V$ minus NIR colors of SN~2001el were dereddened using the following color excesses: 
 $E(V-J$) $=$ 0.450  mag, $E(V-H$) $=$ 0.499  mag and  $E(V-K$) $=$ 0.530  mag.
 These values were derived through comparison  of the color evolution of
 SN~2001el to that 
 of its low reddened twin SN~2004S,  adopting the Galactic reddening law of  \citet{cardelli89} as modified by \citet{odonnell94} (hereafter referred to as the ``CCM$+$O'' law) with an $R_{V} = 3.1$ for the Galactic reddening correction, and an $R_{V}^{\rm host} = 2.15$ for the host reddening portion \citep[see][for details]{krisciunas07}.

\begin{figure}
\figurenum{9}
\epsscale{1.0}
\plotone{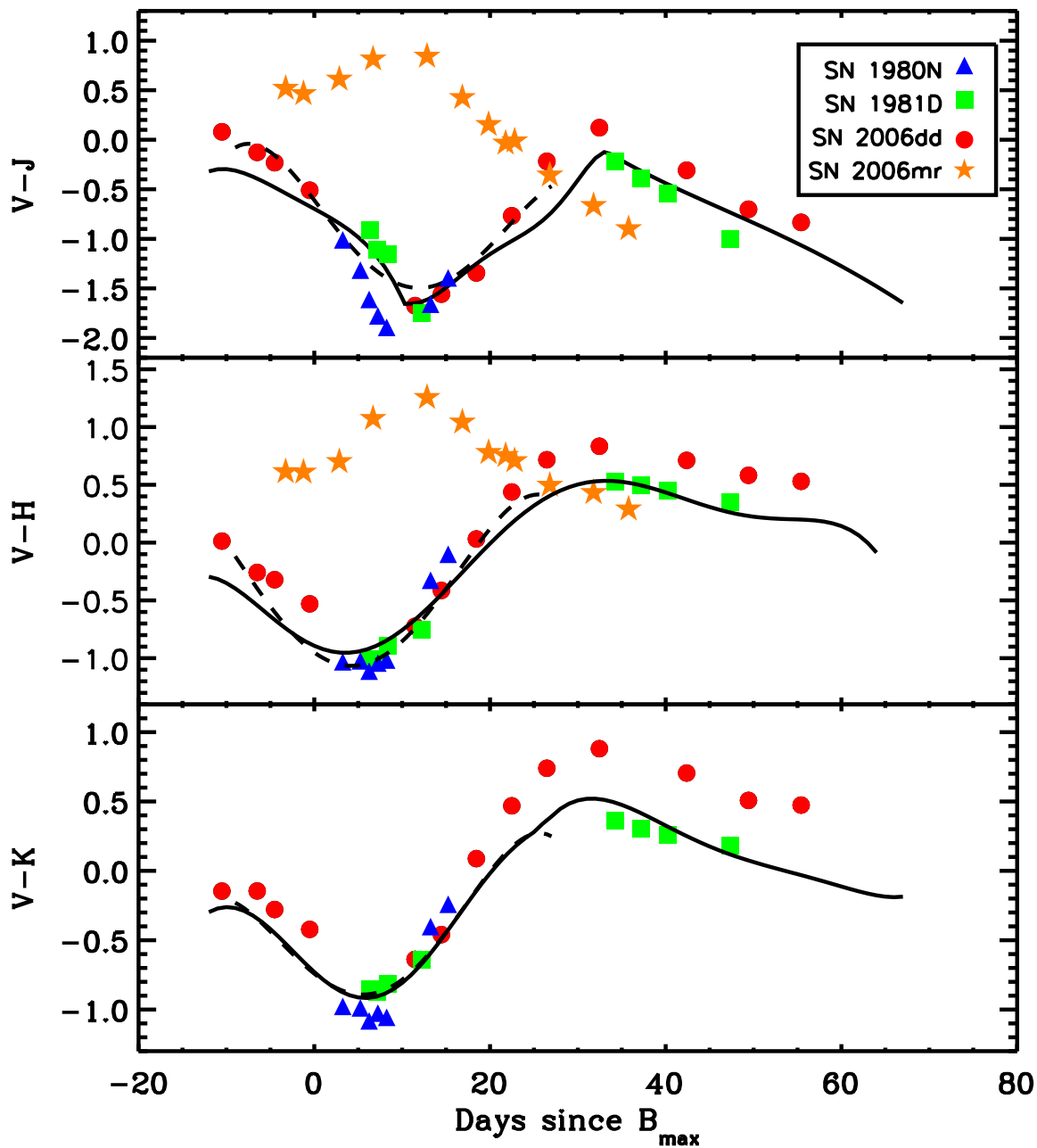}
\caption[Fig9.eps]
{Observed $V$ minus NIR color curves of SN~1980N,
SN~1981D, SN~2006dd, and SN~2006mr.
The data have been $K$-corrected and corrected for Galactic reddening.
Solid lines correspond to zero-reddening loci of  SN~2001el.
For comparison, the loci for midrange decliners is also shown
\citep{krisciunas00}.\label{fig:vjhk}}
\label{fig:ebvhost}
%%{\center Stritzinger {\it et al.} Fig. \ref{fig:FC}}
\end{figure}

Plotted in  Figure~\ref{fig:ebvhost} are the  $(V-J)$, $(V-H)$ and $(V-K)$ color curves of 
 the four SNe in NGC~1316, and the new zero-reddened loci.  Also included for comparison 
 are the loci of  \citet{krisciunas00}. 
When compared to the new loci derived from SN~2001el, each color combination of  SN~1980N and SN~1981D yields slightly negative color excesses, suggesting they suffered essentially no host extinction.
On the other hand, the best fit locus to each 
color curve of SN~2006dd give colors excesses of  
 $E(V-J)  =  0.201~\pm$ 0.169   mag,  
 $E(V-H) =  0.259 ~\pm$ 0.129  mag and 
 $E(V-K) = 0.253 ~\pm$ 0.142   mag, which correspond to  a weighted average reddening of 
$E(B-V) = 0.093~\pm$ 0.038   mag.
When converting the $V$ minus NIR color excesses to $E(B-V)$, we used reddening law 
 coefficients from  Table~13 (columns 4 and 5) of \citet{folatelli10}, along with Equation~A4 of \citet{krisciunas06} and  an $R_{V}^{\rm host} = 3.2~\pm~0.31$.
 This value of  $R_{V}^{\rm host}$ is the 
 favored value   obtained through the comparison of 
 $V$ minus NIR color excess to $E(B-V$)  of SNe~Ia with low-to-moderate reddening
 values \citep[see][Figure 13]{folatelli10}.
The specific conversion formulae are 
$E(B-V)$ $=$ 0.435~$\cdot$~$E(V-J)$,
$E(B-V$) $=$ 0.382~$\cdot$~$E(V-H$), and 
$E(B-V$) = 0.354~$\cdot$~$E(V-K$).
If we instead adopt an $R_{V}^{\rm host} =  1.7~\pm~0.05$, corresponding to the 
reddening law value derived for a sample of moderate-to-low and highly reddened SNe~Ia
\citep[see][Figure~13]{folatelli10}, 
the $V$ minus NIR color excesses derived from the zero-reddened loci of SN~2001el
imply a weighted average reddening of $E(B-V) = 0.156~\pm~0.057$  mag.

Recently, \citet{krisciunas09} noted that the $(V-H)$ and $(V-K)$ colors of 
fast declining SNe~Ia are also remarkably uniform after t $\sim$ +30 days, and 
very similar to the unreddened colors of mid-range decliners.
As shown in Figure~\ref{fig:ebvhost}, the $(V-H)$ color evolution of SN~2006mr
is much redder than that of SN~2006dd until
t $\sim$ +25 days, highlighting the intrinsically red nature of the 
fast-declining SNe~Ia.  After t $\sim$ +25 days, the colors 
of the normal SNe are similar, although not identical.  As NIR light curves of more
fast-declining SNe~Ia are obtained, it will be interesting to see if a 
``NIR Lira law" exists which can be used to derived host galaxy 
reddening over the entire observed range of decline rates of
SNe~Ia.

In summary, we find 
(i)  the EBV method and the Lira relation indicate that SN~1980N is perhaps slightly reddened
[$E(B-V)_{\rm host} =  0.06$  mag] while the $V$ minus NIR colors suggest no host reddening, 
(ii)  the EBV method  indicates SN~1981D has a color excess of 
$E(B-V)_{\rm host} = 0.08$   mag, while the $V$ minus NIR colors indicate
no reddening,  
(iii)  the EBV method and the NIR colors show SN~2006dd suffered a small amount 
of reddening in the range $E(B-V)_{\rm host} \sim $0.04--0.09 mag,
and (iv) SN~2006mr suffered no significant host reddening.
Although the three methods we have employed for estimating the host
reddening are not completely independent --- for example,
the Lira law has historically been used to calibrate the intrinsic colors of SNe~Ia
at maximum light \citep{phillips99,folatelli10} --- all are consistent in indicating that
the four SNe in NGC~1316 were, at most, only slightly reddened.

\citet{maoz08} attempted to identify the progenitors of SNe 2006dd and 2006mr using 
archival {\em Hubble Space Telescope} (HST) images of NGC~1316.
Although they were unable to identify conclusively any point sources, they did
estimate the maximum surface brightness attenuation at the position of each 
SN.  Although there are obvious dust features near both SNe
(see Figure~\ref{fig:FC}), \citet{maoz08}
concluded from an analysis based on iso-brightness contours that the maximum
surface brightness attenuation at the positions of both SNe was
 $A_{V}^{\rm host} = 0.25$~mag, which is consistent with the low host reddening
values we derive from the SNe light curves.  Nevertheless, this method 
is limited by the resolution of the HST images which, while excellent, may not be 
sufficient to resolve the smallest dust filaments or holes located in the central
regions of NGC~1316.
 
\subsection{Distances}
\label{distances}

\subsubsection{SNooPy EBV Method}
\label{snpymethod}
The distance moduli  for SN~1980N, 1981D and 2006dd calculated using
the EBV option in SNooPy (see Table~\ref{tab7}) are consistent,
showing a total variation of only  0.085~mag.   
Combining the distance estimates into a weighted average must be done with care as there are
errors that correlate when averaging over different filters, and errors that correlate 
when averaging over different SN.  

When computing the error in a weighted average of distances estimated from several SNe, one must
consider  the errors introduced by different sources including:  (1) the statistical errors in the best-fit value of the distance from each SN which depend on the quality of the observations, (2) the intrinsic dispersion in SNe~Ia as standard candles, and (3) the uncertainty in the calibration parameters (M$_X$(0), b$_{X}$, 
R${_X}$).  Some of these errors are highly correlated and this must be taken into account.  For instance, an error of $\delta m$ in M$_X(0)$ results in the same systematic error -- $\delta m$ -- in each distance modulus and therefore in the final weighted average.  On the other hand, the intrinsic dispersion in SNe~Ia is a random error and should diminish as 1 / $\sqrt[]{N}$.
These errors can be propagated analytically by building a correlation matrix,
 however, we opted to instead use Monte-Carlo simulations.  
First, the errors listed in Table 9 of \citet{folatelli10} and an intrinsic dispersion in SN~Ia of 0.06 mag was
used to randomly generate 500 realizations of the calibration parameters.  
 We then fit each SN with these realizations, generating 500 distances for each SN.
  This allowed 
 the systematic error in the distance of each event to be determined,  as well as the systematic error in the final weighted average.  From this experiment, the systematic error budget for an individual 
SN is found to be $\approx$ $\pm$0.07 (see Table~\ref{tab7}), whereas for the weighted average distance modulus this reduces to $\pm$0.05. Note that these values  do not include the systematic error in the Hubble constant. 

Combining the SNooPy EBV option average distance moduli listed in Table~\ref{tab7},
we obtain a final weighted average distance modulus 
($\mu_0 - 5\log_{10} \cdot h_{72}$) of
 31.180$\pm$~0.013 (random) $\pm$~0.050 (systematic) mag. 

\subsubsection{Tripp Method}

As discussed above, SN~2006mr was excluded from the SNooPy analysis 
because it is a fast-declining [$\dm = 1.820~\pm~0.020$] event that lies outside of the
range in which the luminosity--decline rate relations of \citet{folatelli10} are applicable. 
However, these authors have suggested that it may be possible 
to compute reliable distances 
from fast-declining SNe~Ia using the two-parameter model 
of \citet{tripp98}.  Following this line of reasoning, we have applied the Tripp method to  
SN~2006mr, as well as to the three normal events.
The Tripp model assumes that the distance modulus  of a SN~Ia
has the following dependence on decline rate and color:   
 
\begin{equation}
\mu_\circ = m_{X}^{\rm max}  -  M_{X}(0)  - b_X \cdot[\dm - 1.1] -          
\beta^{YZ}_{X} \cdot (Y-Z).
\label{beqn7}
\end{equation}

\noindent Here $m_{X}^{\rm max}$ is the observed K-corrected magnitude at maximum
corrected for Galactic reddening, 
M$_{X}^{0}$ is the peak absolute magnitude of SNe~Ia with $\dm = 1.1$ and 
zero color, $b{_X}$ is the slope of the luminosity vs. decline rate relation, 
$\beta^{YZ}_{X}$ is the slope of the luminosity-color relationship, and 
($Y - Z$) is a pseudo-color at maximum.\footnote[12]{A {\em pseudo-color} is 
defined to be the difference between  peak magnitudes of two passbands. 
In the case of SNe~Ia, the time of peak brightness varies from passband to 
passband.}
  
To use the \citet{tripp98} model to estimate distances  requires an accurate 
calibration between the relations of  peak  absolute magnitude vs. $\dm$ 
and pseudo-colors.  These calibrations were obtained using 
25 of the 35 {\em best-observed} SN~Ia light curves presented in the 
first CSP SNe~Ia data release \citep{contreras10}.
The \citet{folatelli10}  calibration of these relations included SN~2006mr.
In order to avoid any circularity we  re-computed the relations omitting this SN. 
The resulting fits of peak absolute 
magnitude vs. $\dm$ and pseudo-color for all of the relevant passbands are
provided in  Table~\ref{tab8}.  Columns 4--6 give the fit values of M$_{X}^{0}$,
$b{_X}$, and $\beta^{YZ}_{X}$ for each filter and pseudo-color combination.
Column 7 gives the values of $R_V$ implied by the slopes of the 
luminosity--color relation, $\beta^{YZ}_{X}$ assuming the CCM$+$O
reddening law, all of which are consistent with a value of $R_V^{\rm host} \sim 1.4$.
With these relations and the measurements 
of $m_{X}^{\rm max}$ and $\dm$ given in Table~\ref{tab5}, 
distance moduli were computed for each observed passband.

Considerable care was also taken to compute
the systematic uncertainty associated with the Tripp distance estimates. 
To do so we performed Monte-Carlo simulations in the same way as in $\S$~\ref{snpymethod}.
Taking the K-corrected peak magnitude of each light curve,
and the calibration parameters M$_{0}$, b$_{X}$,  $\beta^{YZ}_{X}$ and R$_{X}$,
we computed 500 realizations using the
errors of these parameters reported in Table~\ref{tab8},
and adopted the assumption that only M$_{0}$ and R$_{X}$ are  correlated.
From this experiment, we derived a distance modulus for each 
filter combination and a weighted average using the full correlation matrix from the SNooPy fits.
Histograms of the distance moduli were then constructed for each SN 
and the combined average of the three normal events,  from which, the
systematic uncertainty was taken as the  standard deviation of each distribution.

The results of the Tripp method are given in the penultimate column 
of Table~\ref{tab7}.  For  SNe~1980N, 1981D, and 2006dd, the derived average
distance moduli exhibiting a total range of  0.17~mag. 
The weighted average distance modulus ($\mu_0 - 5\log_{10} \cdot h_{72}$)
of these three normal events is
 31.248~$\pm$~0.034 (random) $\pm$~0.040 (systematic) mag.  

In the case of SN~2006mr, the Tripp method yields an average distance modulus
($\mu_0 - 5\log_{10} \cdot h_{72}$) of 
31.834~$\pm$~0.070~$\pm$~0.080 mag, which is more than 0.5~mag greater
than the weighted average of the three normal events.
Contrary to the suggestion of \citet{folatelli10} that fast declining SNe~Ia 
follow the Tripp model, the discrepancy in the distance derived here for SN~2006mr 
compared to the normal objects indicates otherwise.  \citet{folatelli10}  assumed a 
 Surface Brightness  Fluctuations (SBF) derived
distance modulus to NGC~1316 that is $\approx$0.3 mag further than our weighted average, 
and obtained residuals  of $\sim$0.2 mag --- towards a shorter distance --- in all bands with respect 
to the Tripp fits. These differences account for the $\sim0.5$ mag difference obtained here. 

\subsubsection{NIR Method}

The NIR light curves of SNe~1980N, 1981D, 2006dd, and 2006mr
offer an alternative method for deriving the distance to NGC~1316. 
\citet{krisciunas04} showed that there is little or no dependence on 
decline rate of the NIR luminosities of SNe~Ia, and that these objects
are essentially perfect standard candles at these wavelengths.  More
recently, \citet{krisciunas09} have suggested that there may actually be a 
bimodal distribution of SN~Ia luminosities in the NIR in the sense that those 
that peak in the NIR {\em prior} to T($B$)$_{\rm max}$ (like SNe~1980N, 1981D,
and 2006dd) have effectively the same NIR absolute magnitudes at maximum 
light regardless of the value of $\dm$, whereas those that peak in the NIR {\em after} 
T($B$)$_{\rm max}$ (like SN~2006mr) are subluminous in all bands. 

Table~\ref{tab9} gives the average NIR peak absolute magnitudes, M$_{J}$,   M$_{H}$,
and  M$_{K_s}$,  reported in \citet{krisciunas09} but modified to exclude SNe 1980N, 1981D 
and 2006mr.  These values can be used with the reddening values listed in Table~\ref{tab6}
and the observed peak $JHK_s$ magnitudes given in Table~\ref{tab5} to derive independent
distance moduli for NGC~1316.  Combining the distances obtained from the NIR 
light curves of each SNe, we obtain the distance moduli given in the final column
of Table~\ref{tab7}.  Here the  systematic uncertainties were obtained by simply 
adding the errors of M$_{J}$,   M$_{H}$, and  M$_{K_s}$ in quadrature.  
This is due to the fact that \citet{krisciunas09} calibrated their SNe~Ia assuming 
a fixed value of $R_V$ and no dependence of the absolute magnitude on $\dm$.
Therefore, assuming there are no covariances between  M$_{J}$,   M$_{H}$ and  M$_{K_s}$, the systematic error is simply the effect of three independent errors. This also applies for the 
weighted average of the distances derived for the three normal events, for which we find
$\mu_0 - 5.0\log_{10} \cdot h_{72}$ =  31.203~$\pm$~0.012 (random) $\pm$~0.055 (systematic) mag.

The NIR distance moduli computed for the normal objects show a larger total range
(0.24~mag) than those obtained using either the SNooPy EBV or Tripp methods.
This is not surprising since only one of the three, SN~2006dd, was actually observed
at maximum in the NIR.  The weighted average distance modulus of the three
normal SNe is consistent within the quoted uncertainties to the values obtained from 
the EBV and Tripp methods.  When comparing the NIR-determined 
distances to those derived with the SNooPy EBV method, one needs to take into 
account that there is significant covariance between the adopted 
value of R${_V}$ and the derived $JHK_s$ absolute magnitudes listed in 
Table~\ref{tab9}.  Because the \citet{krisciunas09} sample used to derive the values 
listed in  Table~\ref{tab9} has a non-negligible average extinction ($<E(B-V)> = 0.11$),
changing the assumed value of R${_V}$  from 3.1 to 2.0 will change the calibration 
in the following way:

\begin{equation}
\Delta M_J = (R_J[R_V = 3.1] - R_J[R_V = 2.0])\;\cdot<E(B-V)> = 0.05\;{\rm mag}
\label{beqn7.5}
\end{equation}

\begin{equation}
\Delta M_H = (R_H[R_V = 3.1] - R_H[R_V = 2.0])\;\cdot<E(B-V)> = 0.03\;{\rm mag}
\label{beqn8}
\end{equation}

\begin{equation}
\Delta M_K = (R_K[R_V = 3.1] - R_K[R_V = 2.0])\;\cdot<E(B-V)> = 0.02\;{\rm mag}.
\label{beqn9}
\end{equation}

\noindent These differences are in the sense that $\Delta M_X$ should be subtracted from 
the NIR distance  moduli derived using filter $X$.  While \citet{krisciunas09} did not strictly 
assume $R_V=3.1$, these numbers give a rough idea of the magnitude of the correction.

In the case of the fast-declining SN~2006mr, the NIR distance modulus we derive is
$\sim$0.5 mag greater than the normal objects.  Intriguingly, this is in the same sense
and of similar magnitude to the discrepancy found between the distance of 
SN~2006mr and the three normal SNe derived via the Tripp method.
 
 \subsubsection{MLCS2k2}
 
  To ensure that the EBV or Tripp methods, or their adopted calibrations,
  are not the source of the disagreement between the distances derived from the normal
  objects and SN~2006mr, the optical light curves of each SN were also fit with 
 MLCS2k2 \citep{jha07}.  For this purpose we made use  of SN analysis software package 
 {\tt SNANA} \citep{kessler09}.  Optical light curve fits were performed with photometry
 in the {\em standard} Landolt and Smith photometric systems, and  R$_{V}$ was set to be consistent with the calibration used with the EBV method. From this analysis we 
 find a distance to SN~2006mr that is  $\sim$ 50\%  further than the average distance derived from the normal events. This discrepancy is $\sim$ 15\% worse than what was determined from the EBV and Tripp methods.

\section{Na~I~D Absorption}

\begin{figure}
\figurenum{10}
\epsscale{1.0}
\plotone{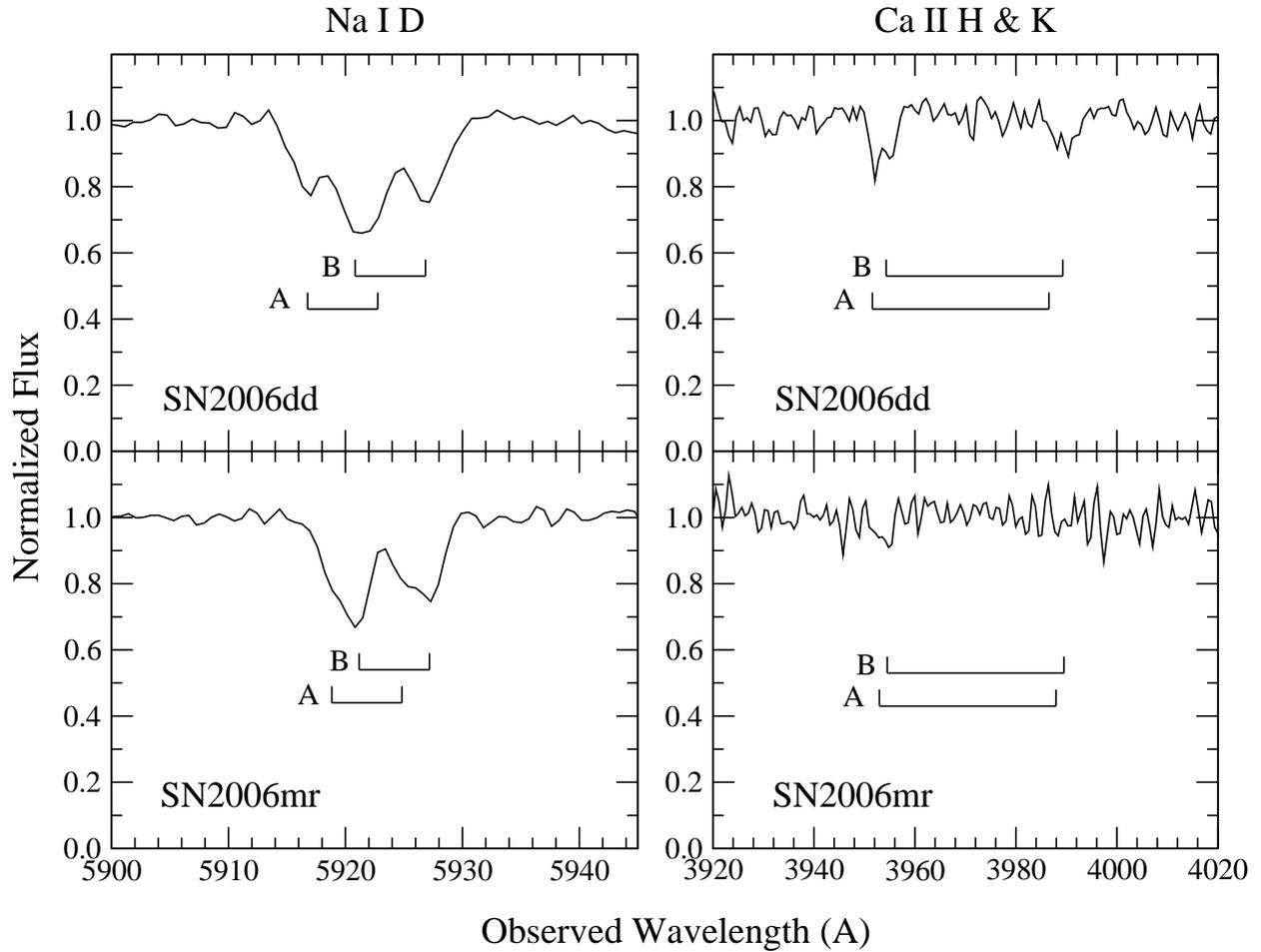}
\caption[Fig10.eps]
{Spectra of SNe~2006dd and 2006mr in the region of the interstellar
Na~I~D and Ca~II~H~\&~K lines in the rest system of NGC~1316.
Indicated by the letters ``A'' and ``B'' are the two strong systems observed in both
SNe.  The spectrum
of SN~2006dd was obtained on 21 June 2006 with a FWHM resolution
of 3.3~\AA; that of SN~2006mr was acquired on 09 Nov. 2006 at a
resolution of 2.4~\AA.\label{fig:NaID}}
%%{\center Stritzinger {\it et al.} Fig. \ref{fig:FC}}
\end{figure}

The spectra of both SNe~2006dd and 2006mr exhibited strong interstellar 
Na~I~D absorption lines in the rest system of the host galaxy.  The left two panels
of Figure~\ref{fig:NaID} show a blow-up of these features in
our highest dispersion spectra.  The total equivalent widths
of the absorption are 3.2~\AA~for SN~2006dd and 2.4~\AA~for SN~2006mr,
although in both cases the profiles show clear evidence for structure.  
It is possible to obtain good fits to the profiles assuming that they are 
composed of two unresolved Na~I~D doublets.  We employed the 
IRAF\footnote[13]{IRAF is distributed by the National Optical Astronomy 
Observatory, which is operated by the Association of Universities for 
Research in Astronomy (AURA) under cooperative agreement with the 
National Science Foundation.}
``fitprofs'' task to deblend the profiles using Gaussian profiles with FWHM
values equal to the spectral resolution.  The equivalent widths derived from 
these fits are given in Table~\ref{tab10}. 
According to \citet{munari97}, the ratio of the D2/D1 equivalent widths is 
2.0 in unsaturated lines, and asymptotically approaches a value of 1.1 at 
high optical depths.  The values of this ratio for the two systems in 2006dd are 
1.2 and 1.3, while in 2006mr the ratios are 1.2 and 1.1. Hence, we are dealing
with saturated lines as expected for such strong absorption.

The right hand panels of Figure~\ref{fig:NaID} show the same spectra of
SNe~2006dd and 2006mr in the region of the Ca~II~H~\&~K lines.
Absorption produced by both H~\&~K is clearly present in SN 2006dd.  In SN 2006mr, the K~line may
be weakly present, but the H~line is hidden by noise.  The total equivalent width of 
the K line in 2006dd is $\sim$0.6~\AA; in 2006mr, it is $\sim$0.2--0.3~\AA.  Hence,
the Ca~II absorption in both SNe is weaker than the Na~I~D absorption.  
A Ca~II/Na~I ratio $\leq 1$ is typical of lines of sight in
the disk of the Milky Way, whereas gas in the halo regions generally shows
Ca~II/Na~I ratio $>> 1$ \citep{cohen75}.  The low ratio in the disk gas is
generally taken as an indication that there has been significant depletion of 
Ca in dust grain formation.

According to NED, the radial velocity of NGC~1316 is 1760~km~s$^{-1}$.
Hence, the interstellar absorption systems observed in {\em both} SNe are
blueshifted with respect to the systemic velocity.  Interestingly, the measured
velocities of the ``B'' systems in both SNe are identical to within the errors.
The ionized gas in NGC~1316 was shown by \citet{schweizer80} to 
exhibit rapid rotation with the maximum velocity gradient running 
approximately NW to SE.  Relative to the nucleus, the SE side is
approaching, and the NW side receding.  SNe~2006dd and 2006mr 
were located to the N and E of the nucleus, respectively.  It therefore
seems unlikely that simple rotation can explain the  
observed blueshifts of the interstellar lines in both SNe unless the gas
producing the absorption has a different origin than the ionized gas.

\begin{figure}[]
 \figurenum{11}
  \epsscale{1.0}
\plotone{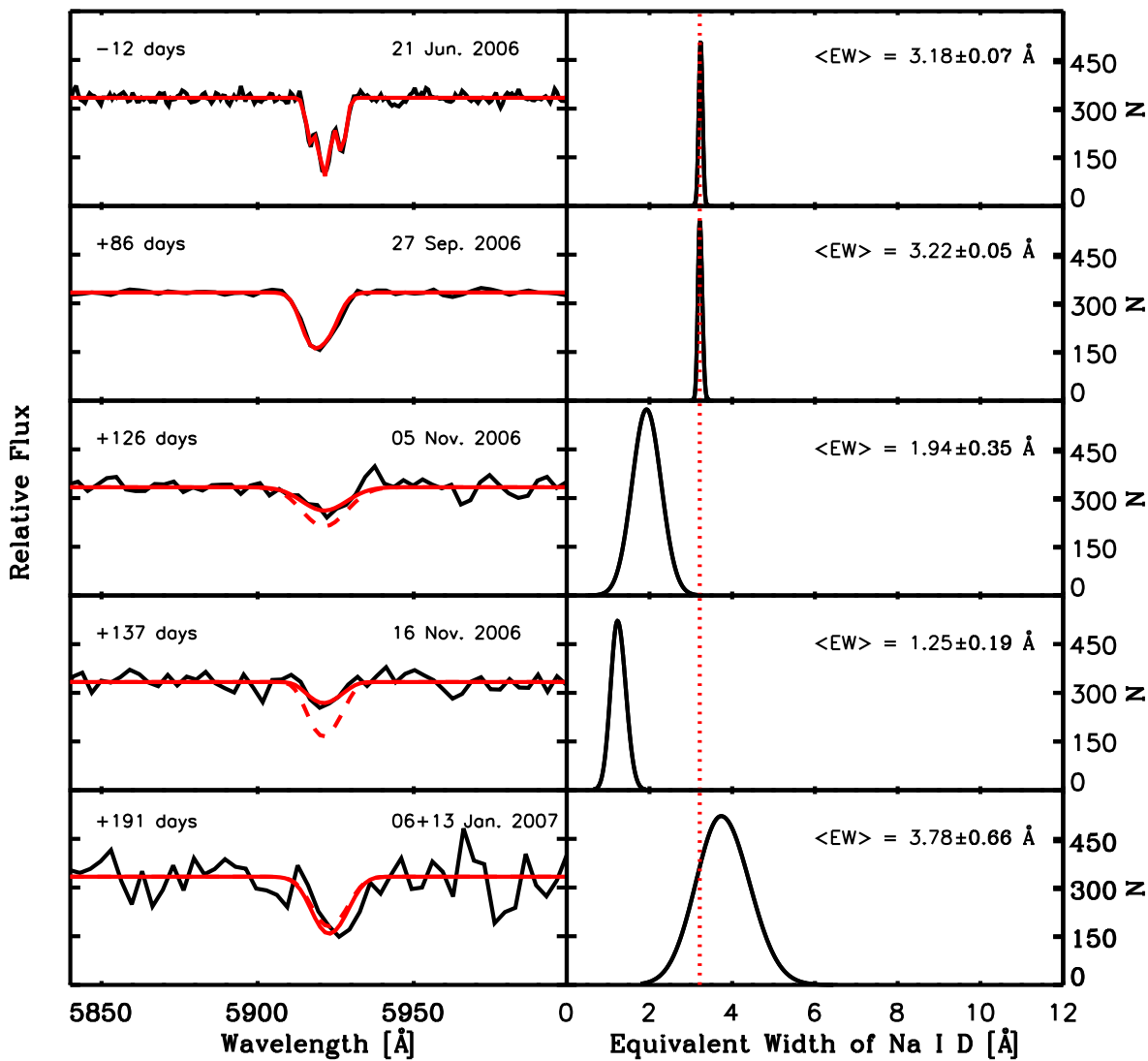}
  \caption[Fig11.eps]
{ ({\em left}) Evolution of the Na~I~D interstellar lines in SN~2006dd  and 
({\em right}) posterior probability distributions for the total equivalent width values
derived from Markov chain Monte Carlo fitting simulations.
The best modeled spectra (red solid lines) are overplotted on
the observed spectra (black solid lines). In the case of the two spectra obtained on 
 05~Nov.~2006 and 16~Nov.~2006, we also include  the model with a total equivalent width of 
 3.2 \AA\ (red dashed lines).
 The final spectrum is actually the sum of two observations made within a week of each other in
Jan.~2007.  To facilitate comparison  of  the PPDs, a vertical (red dotted) line has been drawn at 
3.2~\AA\ to guide the eye. 
\label{fig:NaIevolution}}
  %{\center Stritzinger {\it et al.} Fig. \ref{fig:flcurves}}
\end{figure}

The brightness of SN~2006dd allowed spectral
observations to extend for nearly 7~months. 
Surprisingly, the strength of the absorption appears to have decreased
significantly in the two November 2006 spectra.    
This is highlighted in the  left panel of Figure~\ref{fig:NaIevolution}, which 
displays the evolution of the Na~I~D spectral region from $-14$ to $+$191 days past 
maximum.
In each instance the observed spectrum is over-plotted with the best model determined through Markov chain Monte Carlo (MCMC) fitting simulations.

MCMC is a Bayesian method for determining the posterior probability distribution (PPD) of parameters in a model.  It does this by randomly walking through parameter space, creating a Markov chain of states.  At each point in the chain, the probability of the state is computed by creating a theoretical realization of the data from the model and noise characteristics we define and comparing this with the actual data.  In this way, MCMC is a Monte Carlo simulation.  However, because it is also a Markov Chain, in the sense that each state in the chain is probabilistically determined from the previous state, sampling from this chain after it has converged is equivalent to sampling from the PPD.  MCMC therefore gives us both the maximum likelihood solution to the problem, as well as a robust estimate of the errors and covariances of the parameters.
However, the error estimates depend on the assumed error model of the spectrum.  In all cases, we use Poisson noise based on the observed counts and the read noise of the detector.   It is possible we are neglecting small sources of error (e.g., extra detector noise) and therefore underestimating the uncertainties in the equivalent widths.  This is particularly relevant for our final observation of SN~2006dd, where the signal-to-noise is very low and the significance of the detection, or lack thereof, depends critically on our noise model.

To fit the lines we first modeled the highest resolution, highest signal-to-noise ratio
spectrum (21 June 2006) as two systems, i.e. Na~I~D(A) and Na~I~D(B)  (four Gaussians). 
The separation of the  two doublets was fixed at 5.97 \AA \ while the redshift of each system was allowed to float. 
From this analysis we computed line ratios of 1.18 $\pm$ 0.08 and 1.36 $\pm$ 0.19 and 
total equivalent widths of  1.24 $\pm$ 0.08 \AA\  and 1.94 $\pm$ 0.08 \AA\ for the Na~I~D(A) and Na~I~D(B) doublets,  and  a FWHM $=$ 3.3 \AA.
Note these values are in agreement with what was obtained 
using the  IRAF ``fitprofs'' task (see Table~\ref{tab10}). 
For the remaining spectra the two Na~I doublets are blended together. In this case we imposed priors on all 
the variables consistent with what was obtained  with the 21 June 2006 spectrum, except for 
the two equivalent width components  which were given uniform priors, and the FWHM was set at a constant equal to the resolution of the specific spectrum in question. The total equivalent width values and their associated 1-$\sigma$ error 
are  provided in each panel on the right of Figure~\ref{fig:NaIevolution}.
The right panel of Figure~\ref{fig:NaIevolution} shows the PPDs for the total equivalent width for each epoch. To facilitate comparison the PPDs were placed on a single scale so that the
differences in equivalent width and the error from the models shown in the left panel are apparent.  

\begin{figure}
\figurenum{ 12}
\epsscale{1.0}
\plotone{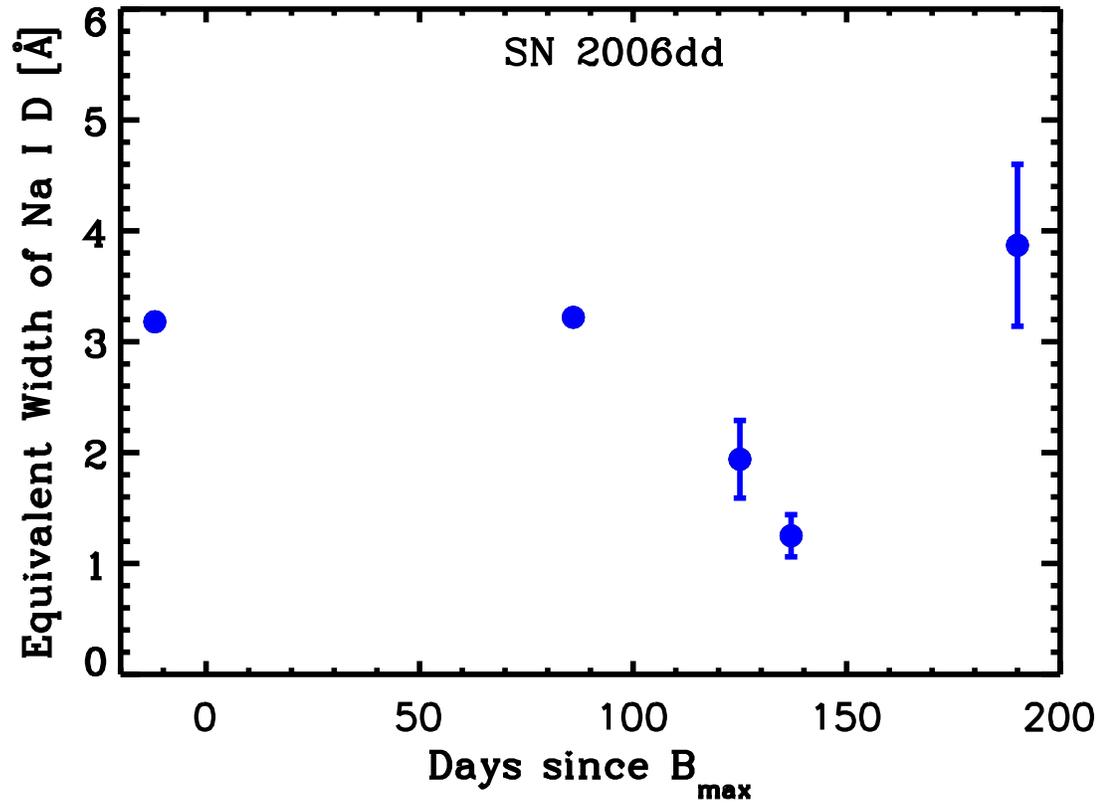}
\caption[Fig12.eps]
{Evolution of the Na~I~D equivalent width in SN~2006dd vs. days past B$_{\rm max}$.
Error-bars are 1-$\sigma$.\label{fig:Naievolution2}}
%%{\center Stritzinger {\it et al.} Fig. \ref{fig:FC}}
\end{figure}

The results of the MCMC  fits are  summarized in Figure~\ref{fig:Naievolution2} where the 
total equivalent width of the Na I D absorption (error bars correspond to 1-$\sigma$) in each spectrum
is plotted vs. days past $B$-band maximum. The weakening in the line strength of Na I D in the two November spectra
is striking.  More quantitatively, we find that the equivalent width of the Nov.~16 line 
differs from the Jun.~21 line at 98\% confidence.  The Nov.~4 equivalent width differs from 
the Jun.~21 value at 93\% confidence.  Taken jointly, the two November equivalent widths differ from the Jun.~21 value at more than 99\% confidence.

Contrary to the situation for SN~2006dd, our spectra 
of SN~2006mr do not show any evidence for a
variation in the strength of the Na~I~D absorption.  However,
our spectroscopic coverage of SN~2006mr extended to only two months
past $B$~maximum, whereas the apparent weakening of the
absorption in SN~2006dd occurred sometime between three and
four and a half months after $B$~maximum.

\begin{figure}
\figurenum{13}
\epsscale{1.0}
\plotone{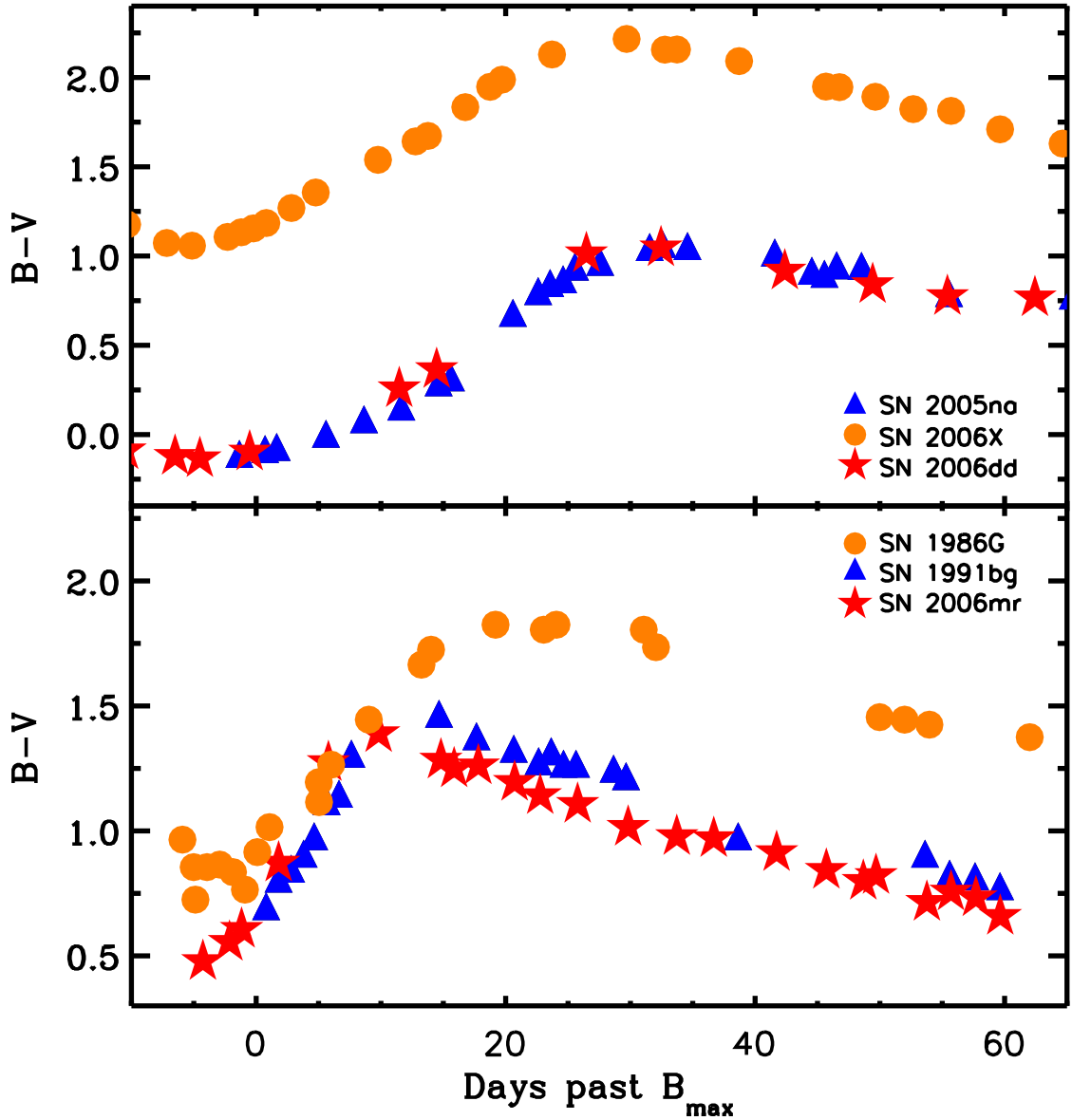}
\caption[Fig13.eps]
{Observed ($B-V)$ color curves of  ({\em top}) normal and ({\em bottom}) low-luminosity SNe~Ia.
% SN~1986G, SN~1991bg, and SN~2006mr. 
Photometry of SN~2005na and SN~2006dd is from \citet{contreras10}, while 
SN~1986G and SN~1991bg originate from \citet{phillips87} and \citet{leibundgut93}, respectively. 
Each color curve has been corrected 
for Galactic reddening.\label{fig:BV}}
%%{\center Stritzinger {\it et al.} Fig. \ref{fig:FC}}
\end{figure}

According to the relation between $E(B-V)$ and
the equivalent width of the Na~I~D1 line  for Galactic stars 
\citep{munari97}, the very strong Na~I~D absorption observed in 
SNe~2006dd and SN~2006mr is totally inconsistent with the
low host galaxy reddening we derive from the light curve observations.
To illustrate the magnitude of this inconsistency, we compare in the upper panel
of Figure~\ref{fig:BV} the ($B-V$) color evolution of SN~2006dd with 
that of two other
SNe~Ia with similar decline rates: SN~2005na, which was essentially unreddened
\citep{folatelli10}, and SN~2006X, which exhibited strong Na~I~D interstellar
absorption with an equivalent width of $\sim$2.0~\AA~and a color excess of 
$E(B-V) \sim 1.2-1.4$  mag \citep{wang08,folatelli10}.  In the lower panel of the
same Figure, the ($B-V$) color evolution of SN~2006mr is compared with
that of two other fast-declining SNe~Ia: SN~1991bg, which appeared in an
elliptical galaxy and is assumed to have been unreddened, and SN~1986G,
which was heavily obscured [$E(B-V) \sim 0.7$  mag] by the dust lane in 
NGC~5128 and showed strong Na~I~D absorption with an equivalent 
width of $\sim3.6$~\AA~\citep{phillips87,phillips99}.  As is seen, the
color evolution of SNe~2006dd and 2006mr closely resembles that
of the two unreddened SNe, yet the strength of the Na~I~D interstellar
absorption was comparable to that observed in the two heavily
reddened SNe.  Clearly the gas-to-dust ratio in  the gas that produced
the Na~I~D absorption in SNe~2006dd and 2006mr is {\em considerably}
different from the typical value in the Milky Way, and also compared to
the gas that produced the strong Na~I~D absorption in SNe~2006X and
1986G.
 
\section{Conclusions}

The four SNe~Ia discovered to date in NGC~1316 provide a unique test of the 
precision of SN~Ia distance determinations.  Employing three different methods,
we find excellent agreement between the three ``normal'' SNe,
1980N, 1981D, and 2006dd for each technique.  The standard deviations
of the distance  moduli for the three SNe using the SNooPy EBV option, the Tripp method, and
NIR light curves are 
  4\%, 8\%, and 10\%, respectively.  Moreover, the agreement
between the average distance  moduli obtained with each technique is exceptional
 (better than 3\%).  On the basis of these results, it
would appear that distances to SNe~Ia can be measured to better than
5\% using present methods.  This finding supports the recent conclusion
of \citet{folatelli10} that it is possible to measure SN~Ia distances to a
relative precision of 3-4\%, as well as the
work of \citet{riess09}, who found a relative dispersion of 4\%
between SNe~Ia distances and Cepheid distances to the host galaxies of 
these same SNe.
 
 Our attempts to derive a distance to NGC~1316 based on the fast-declining
event SN~2006mr were less successful.  Using the two-parameter 
(Tripp) method gives a distance modulus which is $\sim$0.5~mag
larger (or 25 -- 30\% further in distance) than the values derived for the three ``normal'' SNe~Ia
using the same method.  The distance modulus derived for SN~2006mr 
from the NIR light curves is also $\sim$0.5~mag larger than
that obtained for the three normal SNe, even when we use the 
bimodal distribution of absolute magnitudes in the NIR recently
proposed by \citet{krisciunas09}.
Clearly more optical and NIR observations of fast-declining SNe~Ia are
necessary in order to definitively answer the question of whether these
events can be used to determine precise distances.

Based on the three ``normal'' SNe~Ia in NGC~1316, our best estimate 
for the distance modulus of this galaxy is 
($\mu_0 - 5\log_{10} \cdot h_{72}$)  $=$ 31.248~$\pm$~0.034 (random)
$\pm$~0.040 mag (systematic) or 17.8~$\pm$~0.3 (random) $\pm$~0.3 (systematic) 
Mpc for $H_o = 72$.  Here, we have adopted the
distances derived using the Tripp method as representative of
the three methods employed.
This value is in excellent accord with the Planetary 
Nebula Luminosity Function (PNLF) distance modulus of
$\mu = 31.26^{+0.09}_{-0.12}$~mag derived by \citet{feldmeier07}.  
Somewhat worse agreement is found with the SBF measurements of $\mu =  31.59~\pm~0.08$~mag 
by \citet{cantiello07} and $\mu = 31.606~\pm~0.065$~mag by 
\citet{blakeslee09}.  We note that our SN~Ia distance modulus for 
NGC~1316 is $\sim0.4$~mag closer than the value quoted by
\citet{goudfrooij01}, and for which no details were given.  These
authors argued that the SN distance placed NGC~1316 $\sim0.25$~mag
behind the core of the Fornax cluster.   However,
our SN~Ia distance, as well as the PNLF and SBF 
determinations imply that NGC~1316 is very nearly at the same
distance as the cluster core, which SBF observations
place at  $\mu =  31.51\pm0.03$~mag \citep{blakeslee09}. 
A further check is provided by
SN~1992A, which appeared in the Fornax cluster member
NGC~1380.  From photometry presented by \citet{suntzeff96}, we
derive a distance modulus of $\mu =  31.611~\pm~0.008$~mag
using SNooPy and the
absolute magnitude--decline rate calibrations given by
\citet{prieto06}.  The PNLF distance to NGC~1380 is 
$\mu = 31.04^{+0.11}_{-0.12}$~mag, which prompted
\citet{feldmeier07} to speculate that SN~1992A was
$\sim0.4$~mag underluminous.  However, the most
recent SBF determination of $\mu =  31.632~\pm~0.075$~mag
\citep{blakeslee09} is in comfortable agreement with the SN~Ia determination.

The host galaxy reddening values derived from the light curve observations 
for all four SNe in NGC~1316 are modest or negligible.  Nevertheless,
spectra of the two innermost events, 2006dd and 2006mr, show
strong Na~I~D interstellar absorption lines at the redshift of NGC~1316.
Such prominent absorption is {\em always} accompanied by strong
dust reddening in the interstellar medium of our own Galaxy
\citep{munari97}.  The implication is that the
gas-to-dust ratio in the central regions of NGC~1316 is unusually
large.  Although a plausible explanation for this might be that the
dust has been destroyed by UV emission produced from the
relatively recent star formation in NGC~1316 
\citep{schweizer80,goudfrooij01}, the HST images 
published by \citet{maoz08} undeniably reveal the existence
of significant dust near the positions of both SNe.

 If the gas producing the Na~I absorption were to lie inside the dust evaporation
radius of  $\sim10^{17}$~cm \citep{chugai08}, then the high gas-to-dust ratio could be
due to the supernova explosion itself.  The weakening of the Na~I absorption 
at  $\sim$105~days after explosion could, in turn, be due to the SN ejecta overtaking
the cloud producing the absorption.   If we assume a maximum velocity of the ejecta 
of $\sim3\times10^{9}$~cm~s$^{-1}$, then a distance of the gas from the SN of  
$\sim3\times10^{16}$~cm ($\sim$0.01~pc) is implied.  Nevertheless, the flux of a 
SN~Ia is sufficient to ionize gas out to $\sim5\times10^{18}$~cm  \citep{patat07}, so 
the sodium in such a cloud would be ionized at outburst.  The fact that we 
observe a constant equivalent width of the Na~I~D lines from -11.5 to +86.4 days with 
respect to $B$ maximum requires that the Na recombine within a week of explosion.
Even if the gas is in the form of dense ($\sim10^{7}$~cm$^{-3}$) filaments, a gas
cloud lying at a distance of 0.01~pc from the SN would not have recombined
before our first spectroscopic observation \citep{simon09}.  Indeed, a distance of
at least $\sim$10~pc is implied.  The apparent recovery of the Na~I absorption 
$\sim$170~days after explosion also appears to be inconsistent with this scenario
although, as mentioned previously, the reality of the detection of the Na~I absorption 
in our last spectra depends considerably on the assumed signal-to-noise of the spectrum.

Alternatively, the variation of the Na~I absorption could be due to the changing
angular size of the SN as it expands \citep{chugai08}.  
 According to \cite{patat10}, the maximum radius of a SN~Ia in the photospheric phase 
 ($t < 100$~days) is $\sim 10^{17}$~cm.  At the end of the photospheric phase, the effective radius of
the SN should decrease rapidly as the ejecta become optically thin.  Moreover,
the Na~I 
absorption happens to lie near the peak of a [\ion{Co}{3}] emission feature 
which was decaying rapidly over the $\sim$25~day interval when the strong 
decrease in the Na~I absorption was observed. 
It is unclear whether this scenario can account for the possible increase in
the Na~I absorption seen in our final spectra, and does not explain the high
gas-to-dust ratio.

SNe~2006dd and SN~2006mr are not the only apparently low-reddened
SNe~Ia to display strong Na~I~D interstellar absorption lines.
Figure~10 of \citet{folatelli10} indicates that the normal [$\dm = 1.02$]
SN~2005bg had a Na~I~D equivalent width of $\sim$ 2.0 \AA; see also Figure 5 of \citet{blondin09} which contains
several objects with relatively strong Na~I~D absorption accompanied with
low to modest host reddening.  Nevertheless,
the reddening for this SN, as estimated from both the colors at maximum light
and the Lira Law, is negligible.  Absorption in the Ca~II~H~\&K lines 
was also visible in SN~2005bg, with the ratio of the Ca~II/Na~I 
equivalent widths $\sim1$.  Unfortunately, spectroscopic observations
of SN~2005bg were obtained only for a few days around maximum
light, so it is not possible to look for variations of the Na~I~D 
equivalent width.  Indeed, very few SNe~Ia have been observed
with signal-to-noise ratio large enough to be 
able to detect a variation like that observed in SN~2006dd.

The glaring inconsistency between the presence of strong
interstellar line absorption, yet the absence of evidence for
significant dust reddening highlights our current ignorance of the
host galaxy reddening properties of SNe~Ia.  On the one hand,
we have shown in this paper that application of three different
methods for determining distances to SNe~Ia gives results that
are stunningly consistent for the three normal SNe~Ia in
NGC~1316, and which support recent evidence
that these objects can be used to measure cosmological 
distances to a precision of $< 5$\%.  On the other hand, we
are at a loss to explain how two of the SNe in NGC~1316
could show such strong interstellar Na~I~D lines when 
the same methods used to determine such precise distances
indicate that both SNe suffered small or negligible host
galaxy reddening.  The treatment of dust extinction in SNe~Ia
is considered by many \citep[e.g., see ][]{wood-vasey07}
to be the single largest source of systematic 
error in using these objects to constrain the dark energy 
equation of state in next-generation experiments.  While the
SNe in NGC~1316  confirm that SNe~Ia are potentially
one of the most powerful tools
for studying cosmology, they also remind us
that there is still much left to understand regarding the
properties of these explosions and their progenitor 
systems.

\acknowledgments 

We owe a debt of gratitude to Rick Kessler for performing the 
MLCS2k2 fits discussed in $\S$~3.3.4.
We are  grateful to D. Watson, J. Prieto,  P. H\"oflich, J. Sollerman and  J. Fynbo 
for stimulating discussions related to this work.
We acknowledge  the Aspen Center for Physics for hosting the summer meeting ``Taking Supernova Cosmology into the Next Decade", during which this paper was being finalized and benefitted from discussions with other participants.
A special thanks to J. Prieto, D. DePoy, L. Watson,  C. Morgan and  M. Eyler for the 05 Nov. 2006 spectrum of SN~2006dd, and 
the 20 Nov. 2006 spectrum of SN~2006mr.
This material is based upon work supported by the National Science
Foundation (NSF) under grant AST--0306969. 
The Dark Cosmology Centre is funded by the Danish NSF. 
GF and MH acknowledges support from Iniciativa Cientifica Milenio
through grant P06-045-F and CONICYT through Centro de Astrofisica
FONDAP (grant 15010003), Programa Financiamiento Basal (grant PFB 06),
and Fondecyt (grant 1060808 and 3090004).
The CTIO 1.3-m telescope is operated by the Small and Moderate Aperture Research 
Telescope System (SMARTS) Consortium.
We have made use of the NASA/IPAC Extragalactic
Database (NED) which is operated by the Jet Propulsion Laboratory,
California Institute of Technology, under contract with the National
Aeronautics and Space Administration. 
This publication makes use of data products from the Two Micron All Sky Survey, which is a joint project of the University of Massachusetts and the Infrared Processing and Analysis Center/California Institute of Technology, funded by the National Aeronautics and Space Administration and the National Science Foundation.

\clearpage
\begin{deluxetable} {lclcclccc}
\tabletypesize{\scriptsize}
\tablecolumns{9}
%\tablewidth{0pt}
\tablenum{1}
\tablecaption{Spectroscopic observations\label{tab1}}
\tablehead{
\colhead{Date} &
\colhead{Julian Date} &
\colhead{Epoch\tablenotemark{a}} &
\colhead{Telescope} &
\colhead{Instrument} &
\colhead{Range} &
\colhead{Resolution} &
\colhead{Ncombine} &
\colhead{Integration} \\ %&
%\colhead{$\langle S/N \rangle$\tablenotemark{c}}\\
\colhead{} &
\colhead{JD$-2,453,000$} &
\colhead{(days)} &
\colhead{} &
\colhead{} &
\colhead{(\AA)} &
\colhead{(FWHM \AA)} &
\colhead{} &
\colhead{(sec)}}% &
%\colhead{(in 10 \AA)}}
\startdata
\multicolumn{8}{c}{\bf 2006dd}\\
2006 June 21&907.9    &  $-$11.5      & Baade   & IMACS  & 3410 -- 9610  & 3.3  & 2 & 600 \\ 
2006 Sept. 27&1005.8   &  $+$86.4     & Du~Pont & WFCCD  & 3800 -- 9235  & 8.0  & 3 & 600 \\ %sept 27
2006 Nov. 5  &1046.5   & $+$125.5     & Hiltner & B\&C   & 3868 -- 7320  & 14  & 4 & 600 \\
2006 Nov. 16&1055.7   &  $+$136.8     & Du~Pont & WFCCD  & 3800 -- 9235  & 8.0  & 3 & 600 \\
2007 Jan. 6&1106.8   &  $+$187.4      & NTT     & EMMI   & 3500 -- 9610  & 8.0  & 3 & 200 \\
2007 Jan. 13&1113.8   &  $+$194.4     & Du~Pont & B\&C   & 3905 -- 10000  & 8.0  & 3 & 900 \\
\hline
\multicolumn{8}{c}{\bf 2006mr}\\
2006 Nov. 9  & 1048.8   &  $-$2.1  & Clay  & LDSS     & 3800 -- 9975   & 2.4 & 2 & 120   \\
2006 Nov. 13 & 1052.8   &  $+$1.9  & Baade & IMACS    & 4260 -- 9510   & 3.6 & 2 & 150   \\ %needs updated
2006 Nov. 16 & 1055.7   &  $+$4.9  & Du~Pont & WFCCD  & 3800 -- 9235   & 8.0 & 3 & 300   \\
2006 Nov. 19 & 1058.5   &  $+$7.6  & Du~Pont & WFCCD  & 3800 -- 9235   & 8.0 & 3 & 300   \\
2006 Nov. 20 & 1059.5   &  $+$8.5  & Hiltner & B\&C   & 3835 -- 7289   & 14 & 3 & 300   \\
2006 Nov. 22 & 1061.8   &  $+$10.8& Du~Pont & WFCCD   & 3800 -- 9235   & 8.0 & 3 & 300   \\
2006 Dec. 18 & 1087.8   &  $+$36.8 & Du~Pont & WFCCD  & 3860 -- 9235   & 8.0 & 3 & 500   \\
2007 Jan. 13 & 1113.8   &  $+$62.9 & Du~Pont & B\&C   & 3905 -- 10000   & 8.0 & 3 & 600    \\
\enddata
\tablenotetext{a}{Days since T($B)_{\rm max}$.} 
\end{deluxetable}

\label{tab1}
\clearpage
\begin{deluxetable}{cccccccccc}
\tabletypesize{\footnotesize}
\tablecolumns{9}
\tablenum{2}
\rotate
\tablewidth{0pt}
\tablecaption{Photometry of the CTIO Local Sequence in the Standard System\label{tab2}}
\tablehead{
\colhead{ID} &
\colhead{R.A.} &
\colhead{Decl.} &
\colhead{$B$} &
\colhead{$V$} &
\colhead{$R$} &
\colhead{$I$} &
\colhead{$J$} & 
\colhead{$H$} & 
\colhead{$K_s$} }
\startdata      
CSP02 & 3$^{\rm h}$22$^{\rm m}$38${\fs}$36 & $-$37$\arcdeg$14$\arcmin$45$\farcs$31 & 16.180(005) &  15.567(007) & 15.203(008) & 14.858(008)  & 12.807(006)  & 12.213(007) & 12.077(012)\\ % KK2       
CSP03 & 3$^{\rm h}$22$^{\rm m}$31${\fs}$62 & $-$37$\arcdeg$13$\arcmin$51$\farcs$46 & 16.812(008) &  15.682(004) & 14.996(004) &  14.412(008)  & $\cdots$  &  $\cdots$  & $\cdots$   \\         %KK3
CSP06 & 3$^{\rm h}$22$^{\rm m}$34${\fs}$23 & $-$37$\arcdeg$10$\arcmin$01$\farcs$71 & 17.470(008) &  16.859(006) & 16.489(007) &  16.114(006)  & $\cdots$  &  $\cdots$  & $\cdots$  \\          %KK4
CSP50 & 3$^{\rm h}$22$^{\rm m}$44${\fs}$55 & $-$37$\arcdeg$12$\arcmin$46$\farcs$94 & 16.204(011) &  15.027(005) & 14.288(007) & 13.684(008)  & $\cdots$  &  $\cdots$  & $\cdots$\\ %KK#1
\enddata
\tablecomments{Uncertainties given in parentheses in millimag.}
%K1 & 50.68520 & -37.21304 
%K2 & 50.65982 
%K3 & 50.63173
%K4 & 50.64264
\end{deluxetable}

\label{tab2}
\clearpage
\begin{deluxetable}{ccccccccc}
\tabletypesize{\scriptsize}
%\rotate
\tablecolumns{9}
\tablenum{3}
\tablewidth{0pt}
\tablecaption{Optical and Near-IR Photometry of SN~2006dd in the natural system of the CTIO 1.3-m telescope\label{tab3}}
\tablehead{
\colhead{JD$-2,453,000$} &
\colhead{Phase\tablenotemark{a}} &
\colhead{$B$} &
\colhead{$V$} &
\colhead{$R$} &
\colhead{$I$} &
\colhead{$J$} & 
\colhead{$H$} & 
\colhead{$K_s$} }
\startdata
908.90 & $-$10.5 & 13.579(009) &   13.629(007) & 13.423(008)  & 13.398(011)  & 13.477(021)& 13.537(024)& 13.692(034)  \\   
912.91 & $-$6.5  & 12.713(008) &   12.788(007) & 12.612(008)  & 12.642(010)  & 12.851(013)& $\cdots$   & 12.853(020)  \\
914.90 & $-$4.5  & 12.498(009) &   12.588(009) & 12.437(008)  & 12.499(009)  & 12.755(019)& 12.843(019)& 12.784(026)  \\
918.90 & $-$0.5  & 12.336(009) &   12.380(008) & 12.290(013)  & 12.516(011)  & 12.825(021)& 12.850(025)& 12.714(028)  \\
930.88 & $+$11.5 & 13.033(017) &   12.733(009) & 12.755(011)  & 13.103(013)  & 14.346(037)& 13.393(025)& 13.259(023)  \\ 
933.88 & $+$14.5 & 13.347(009) &   12.948(008) & 12.902(009)  & 13.161(010)  & 14.444(029)& 13.296(022)& 13.290(023)  \\
937.86 & $+$18.5 &  $\cdots$   &    $\cdots$   & $\cdots$     & $\cdots$     & 14.456(029)& 13.063(021)& 12.974(021) \\
941.89 & $+$22.5 &  $\cdots$   &    $\cdots$   & $\cdots$     & $\cdots$     & 14.139(026)& 12.909(020)& 12.858(020) \\
945.88 & $+$26.5 & 14.682(011) &   13.643(009) & 13.167(010)  & 12.936(011)  & 13.821(023)& 12.856(019)& 12.823(020) \\
951.86 & $+$32.5 & 15.106(015) &   14.039(014) & 13.559(012)  & 13.241(023)  & 13.878(034)& 13.133(025)& 13.082(025) \\
961.79 & $+$42.4 & 15.430(011) &   14.465(008) & 14.064(013)  & 13.827(011)  & 14.725(031)& 13.687(025)& 13.660(023) \\
968.83 & $+$49.5 & 15.547(013) &   14.663(011) & 14.308(016)  & 14.128(019)  & 15.323(040)& 14.024(027)& 14.066(022) \\
974.82 & $+$55.4 & 15.634(011) &   14.816(011) & 14.518(013)  & 14.400(018)  & 15.607(046)& 14.233(031)& 14.253(023) \\
980.84 & $+$61.4 & $\cdots$    &   $\cdots$    & $\cdots$     & $\cdots$     & 16.081(021)\tablenotemark{b}& 14.606(020)\tablenotemark{b}& $\cdots$    \\
993.82 & $+$74.3 & $\cdots$    &   $\cdots$    & $\cdots$     & $\cdots$     & 16.798(043)\tablenotemark{b}& $\cdots$     $\cdots$    \\
995.85 & $+$76.4 & $\cdots$    &   $\cdots$    & $\cdots$     & $\cdots$     & 16.931(035)\tablenotemark{b}& 15.267(020)\tablenotemark{b}& $\cdots$    \\
\enddata
\tablecomments{Uncertainties given in parentheses in millimag 
and include the usual photon statistics as well as errors in the zero points.}
\tablenotetext{a}{Days since T$(B)_{\rm max}$.}
\tablenotetext{b}{Photometry obtained with the Swope telescope equipped with RetroCam. This photometry is in the natural system of the Swope telescope.}
\end{deluxetable}

\label{tab3}
\clearpage
\begin{deluxetable}{cccccccc}
\tabletypesize{\scriptsize}
%\rotate
\tablecolumns{8}
\tablenum{4}
\tablewidth{0pt}
\tablecaption{Optical Photometry of SN~2006dd in the natural system of the Swope telescope\label{tab4}}
\tablehead{
\colhead{JD$-2,453,000$} &
\colhead{Phase\tablenotemark{a}} &
\colhead{$u$} &
\colhead{$g$} &
\colhead{$r$} &
\colhead{$i$} &
\colhead{$B$} & 
\colhead{$V$} }
\startdata
979.92  & $+$60.534 & $\cdots$    & 15.286(004) &  14.804(004) & $\cdots$    & $\cdots$    & $\cdots$    \\
981.84  & $+$62.451 & 16.893(019) & 15.301(006) &  14.829(009) & 15.041(009) & 15.691(007) & 14.903(007) \\  
986.84  & $+$67.452 & 16.997(021) & 15.391(005) &  15.025(004) & 15.236(006) & 15.784(007) & 15.047(006) \\
994.85  & $+$75.462 & 17.154(019) & 15.521(004) &  15.265(006) & 15.514(008) & 15.892(006) & 15.232(006) \\
998.86  & $+$79.469 & 17.263(021) & 15.585(006) &  15.393(007) & 15.654(009) & 15.958(007) & 15.326(007) \\
1001.90 & $+$82.509 & 17.308(022) & 15.632(009) &  15.488(011) & 15.736(016) & 16.017(008) & 15.407(007) \\
1007.91 & $+$88.518 & $\cdots$    & $\cdots$    &   $\cdots$   &  $\cdots$   & 16.102(010) & 15.553(006) \\
1014.82 & $+$95.435 & 17.606(027) & 15.858(011) &  15.916(015) & 16.173(013) & 16.199(007) & 15.717(006) \\
1017.83 & $+$98.438 & 17.712(031) & 15.900(008) &  16.013(009) & 16.277(012) & 16.265(011) & 15.786(012) \\
1046.68 & $+$127.28 & 18.429(038) & 16.373(006) &  16.863(009) & 17.040(013) & 16.731(009) & 16.394(009) \\
1048.82 & $+$129.42 & 18.553(043) & 16.415(006) &  16.910(010) & 17.105(015) & 16.750(009) & 16.436(008) \\
1049.78 & $+$130.39 & 18.518(035) & 16.432(015) &  16.936(023) & 17.112(022) & 16.777(014) & 16.439(013) \\
1052.75 & $+$133.36 & 18.494(032) & 16.504(015) &  17.024(026) & 17.181(027) & 16.817(013) & 16.515(016) \\
1056.74 & $+$137.35 & 18.658(037) & 16.548(009) &  17.142(018) & 17.282(023) & 16.898(011) & 16.578(013) \\
1060.78 & $+$141.38 & 18.677(045) & 16.622(016) &  17.245(024) & 17.371(033) & 16.944(014) & 16.680(018) \\
1065.76 & $+$146.37 & 18.828(043) & 16.713(013) &  17.373(019) & 17.447(028) & 17.024(014) & 16.769(018) \\
1066.83 & $+$147.44 & 18.853(035) & 16.709(011) &  17.384(019) & 17.515(026) & 17.034(014) & 16.777(014) \\
1068.76 & $+$149.37 & 18.906(038) & 16.740(010) &  17.462(022) & 17.517(028) & 17.066(015) & 16.796(016) \\
1071.67 & $+$152.27 & 18.924(051) & 16.789(011) &  17.547(018) & 17.597(033) & 17.155(013) & 16.891(014) \\
1073.73 & $+$154.33 & 18.947(059) & 16.822(009) &  17.577(016) & 17.605(023) & 17.161(011) & 16.908(012) \\
1076.72 & $+$157.33 & 19.079(073) & 16.868(009) &  17.663(018) & 17.667(027) & 17.224(012) & 16.969(012) \\
1080.78 & $+$161.38 & 19.170(059) & 16.938(009) &  17.729(023) & 17.730(032) & 17.279(012) & 17.031(011) \\
1084.68 & $+$165.29 & 19.252(042) & 16.994(009) &  17.824(023) & 17.812(022) & 17.345(012) & 17.100(012) \\
1087.63 & $+$168.24 & 19.300(044) & 16.983(015) &  17.867(030) & 17.830(032) & 17.395(012) & 17.073(018) \\
1092.68 & $+$173.29 & 19.475(045) & 17.050(013) &  17.989(031) & 17.955(031) & 17.483(014) & 17.199(016) \\
1096.66 & $+$177.27 & 19.521(052) & 17.141(009) &  18.099(023) & 18.012(029) & 17.537(011) & 17.262(012) \\
1099.63 & $+$180.24 & 19.520(075) & 17.232(009) &  18.183(020) & 18.063(029) & 17.567(010) & 17.359(012) \\
1100.65 & $+$181.26 & 19.889(086) & 17.259(008) &  18.210(022) & 18.107(028) & 17.605(011) & 17.352(013) \\
1104.72 & $+$185.33 & $\cdots$    & 17.298(012) &  18.257(025) & 18.104(034) & 17.682(018) & 17.442(016) \\
1106.65 & $+$187.26 & $\cdots$    & 17.333(008) &  18.338(025) & 18.146(033) & 17.700(012) & 17.472(014) \\
1108.65 & $+$189.26 & $\cdots$    & 17.303(014) &  18.378(032) & 18.167(042) & 17.731(013) & 17.460(017) \\
1110.61 & $+$191.22 & $\cdots$    & 17.416(015) &  18.380(049) & 18.152(045) & 17.773(019) & 17.548(021) \\
1118.67 & $+$199.28 & $\cdots$    & 17.515(009) &  18.568(038) & 18.351(055) & $\cdots$    & $\cdots$   \\ 
1123.63 & $+$204.24 & $\cdots$    & 17.585(010) &  18.762(040) & 18.471(051) & 17.943(016) & 17.750(017) \\
1127.69 & $+$208.29 & $\cdots$    & 17.646(011) &  18.768(056) & $\cdots$    & 17.988(015) & 17.812(015) \\
1130.57 & $+$211.17 & $\cdots$    & 17.712(013) &  18.917(041) & 18.529(044) & 18.069(041) & 17.835(017) \\
1133.60 & $+$214.21 & $\cdots$    & 17.727(012) &  18.898(038) & 18.492(040) & 18.112(016) & 17.902(017) \\
1136.63 & $+$217.23 & $\cdots$    & 17.773(010) &  19.016(046) & 18.559(051) & 18.124(025) & 17.954(019) \\
1143.56 & $+$224.17 & $\cdots$    & 17.863(016) &  19.117(051) & 18.686(051) & 18.263(018) & 17.995(022) \\
1147.57 & $+$228.17 & $\cdots$    & 17.957(013) &  19.151(056) & 18.753(064) & 18.342(017) & 18.125(020) \\
\enddata
\tablecomments{Uncertainties given in parentheses in millimag 
and include the usual photon statistics as well as errors in the zero points.}
\tablenotetext{a}{Days since T$(B)_{\rm max}$.}
\end{deluxetable}

\label{tab4}
\clearpage
\begin{deluxetable}{lccccccccc}
\tabletypesize{\scriptsize}
\rotate
\tablecolumns{10}
\tablenum{5}
\tablewidth{0pt}
\tablecaption{Maximum Light Magnitudes of SNe~1980N, 1981D, 2006dd \& 2006mr\label{tab5}}
\tablehead{
\colhead{} &
\colhead{T($B$)$_{\rm max}$} &
\colhead{} &
\colhead{} &
\colhead{} &
\colhead{} &
\colhead{} &
\colhead{} &
\colhead{} &
\colhead{} \\
\colhead{SN} &
\colhead{JD-2,400,000} &
\colhead{$\dm$$^{a}$} &
\colhead{$B$} &
\colhead{$V$} &
\colhead{$r$} &
\colhead{$i$} &
\colhead{$J$} &
\colhead{$H$} &
\colhead{$K_s$}}
\startdata

1980N  & 44584.36$\pm$0.24 & 1.105$\pm$0.019 & 12.375$\pm$0.016 & 12.306$\pm$0.016 &  12.361$\pm$0.027 & 12.952$\pm$0.035 & 12.679$\pm$0.062 & 13.035$\pm$0.042 & 13.047$\pm$0.102 \\
1981D  & 44679.51$\pm$0.30 & 1.212$\pm$0.050 & 12.600$\pm$0.051 & 12.381$\pm$0.025& $\cdots$         & $\cdots$         & 12.886$\pm$0.055& 13.047$\pm$0.029 & 13.252$\pm$0.049 \\
2006dd & 53919.41$\pm$0.08 & 1.080$\pm$0.014 & 12.238$\pm$0.009 & 12.311$\pm$0.009 & 12.310$\pm$0.011 & 12.832$\pm$0.010 & 12.731$\pm$0.014 & 12.839$\pm$0.020 & 12.807$\pm$0.060 \\
2006mr & 54050.94$\pm$0.12 & 1.820$\pm$0.020 & 15.402$\pm$0.014 & 14.612$\pm$0.010 & 14.378$\pm$0.014 & 14.503$\pm$0.009 & 14.056$\pm$0.007 & 13.850$\pm$0.006 &  $\cdots$         \\
\enddata
\tablenotetext{a}{Derived from SNooPy ``max" fits or in the case of 
SN~2006mr via spline fits.}
\tablecomments{Peak magnitudes estimated from photometry K-corrected and corrected for Galactic reddening. 
}

\end{deluxetable}

\label{tab5}
\begin{deluxetable}{lccc}
\tabletypesize{\normalsize}
%\tabletypesize{\scriptsize}
%\rotate
\tablecolumns{4}
\tablenum{6}
\tablewidth{0pt}
\tablecaption{Host Galaxy Reddenings of SNe~Ia in NGC~1316\label{tab6}}
\tablehead{
\colhead{} &
\multicolumn{3}{c}{$E(B-V)_{\rm Host}$} \\
\colhead{SN} &
\colhead{SNooPy EBV} &
\colhead{Lira Law} &
\colhead{$V-$NIR}} 
\startdata
1980N  &  0.060~$\pm$~0.011      &  0.081~$\pm$~0.027        & $-$0.021~$\pm$~0.010 \\
1981D  & 0.078~$\pm$~0.024 &  $\cdots$                       & $-$0.025~$\pm$~0.015 \\
2006dd & 0.043~$\pm$~0.008 & $-$0.013~$\pm$~0.005 &  0.093~$\pm$~0.038 \\
2006mr & $\cdots$                      & $-$0.025~$\pm$~0.013            & $\cdots$           \\
\enddata
\end{deluxetable}

\label{tab6}
\clearpage
\begin{deluxetable}{lcccc}
\tabletypesize{\scriptsize}
%\rotate
\tablecolumns{5}
\tablenum{7}
\tablewidth{0pt}
\tablecaption{Distances of SNe~Ia in NGC~1316\label{tab7}}
\tablehead{
\colhead{} &
\colhead{} &
\colhead{$\mu_0$$^b$} & 
\colhead{$\mu_0$$^b$} & 
\colhead{$\mu_0$$^c$} \\
\colhead{SN} &
\colhead{$\dm$$^a$} &
\colhead{SNooPy EBV} &
\colhead{Tripp} &
\colhead{NIR}}
\startdata
1980N   & 1.105~$\pm$~0.011       & 31.211~$\pm$~0.028~$\pm$~0.065 & 31.264~$\pm$~0.055~$\pm$~0.060 & 31.238~$\pm$~0.034~$\pm$~0.055 \\
1981D   & 1.212~$\pm$~0.071 & 31.242~$\pm$~0.036~$\pm$~0.081 & 31.103~$\pm$~0.090~$\pm$~0.080 & 31.398~$\pm$~0.033~$\pm$~0.055 \\
2006dd  & 1.078~$\pm$~0.025 & 31.157~$\pm$~0.016~$\pm$~0.068 & 31.276~$\pm$~0.048~$\pm$~0.070 & 31.162~$\pm$~0.014~$\pm$~0.055 \\
combined& $\cdots$                      & 31.180~$\pm$~0.013~$\pm$~0.050       & 31.248~$\pm$~0.034~$\pm$~0.040 & 31.203~$\pm$~0.012~$\pm$~0.055 \\
\hline
2006mr  & 1.820~$\pm$~0.020 & $\cdots$                       & 31.834~$\pm$~0.070~$\pm$~0.080 & 31.739~$\pm$~0.005~$\pm$~0.052 \\
\enddata
\tablenotetext{a}{Derived from SNooPy EBV fits or in the case of SN~2006mr via spline fit.}
\tablenotetext{b}{Uncertainties correspond to fit and systematic errors. The systematic errors were computed 
from Monte-Carlo simulations (see $\S$~\ref{distances}). Also note these errors do not include the uncertainty associated with the Hubble constant.}
\tablenotetext{c}{Uncertainties correspond to fit, reddening and systematic errors (see text).}
\end{deluxetable}

\label{tab7}
\clearpage
\begin{deluxetable}{lccccccccc} 
\tabletypesize{\scriptsize} 
\tablecolumns{10} 
\tablenum{8}
\tablewidth{0pt} 
%\rotate 
\tablecaption{Fits of peak magnitudes versus $\dm$ and pseudocolor\label{tab8}} 
\tablehead{ 
\colhead{Fit} & \colhead{Filter} & \colhead{Pseudocolor} & \colhead{} & \colhead{} & \colhead{} & \colhead{$R_V$} & \colhead{$\sigma_{\rm SN}$} & \colhead{RMS} & \colhead{} \\ %& \colhead{} \\ 
\colhead{No.} & \colhead{$X$} & \colhead{$(Y-Z)$} & \colhead{$M_X(1.1,0)$} & \colhead{$b_X$} & \colhead{$\beta_X^{YZ}$} & \colhead{(CCM+O)} & \colhead{[mag]} & \colhead{[mag]} & \colhead{$N_{\rm SNe}$} \\ %& \colhead{Sample} \\ 
\colhead{(1)} & \colhead{(2)} & \colhead{(3)} & \colhead{(4)} & \colhead{(5)} & \colhead{(6)} & \colhead{(7)} & \colhead{(8)} & \colhead{(9)} & \colhead{(10)} }
% & \colhead{(11)} }
\startdata
$1$ & $B$ & $(B-V)$ & $-19.09$$\pm$$0.02$ & \phs$0.70$$\pm$$0.10$ & \phs$2.70$$\pm$$0.11$ & \phs$1.42$$\pm$$0.10$ & $0.11$ & $0.15$ & 25 \\ %& best-observed \\ 
$2$ & $V$ & $(B-V)$ & $-19.09$$\pm$$0.02$ & \phs$0.70$$\pm$$0.10$ & \phs$1.70$$\pm$$0.11$ & \phs$1.42$$\pm$$0.10$ & $0.11$ & $0.15$ & 25 \\ %& best-observed \\ 
$3$ & $u$ & $(u-V)$ & $-19.35$$\pm$$0.03$ & \phs$0.52$$\pm$$0.10$ & \phs$1.73$$\pm$$0.04$ & \phs$1.08$$\pm$$0.07$ & $0.09$ & $0.13$ & 24 \\ %& best-observed \\ 
$4$ & $g$ & $(g-r)$ & $-18.85$$\pm$$0.01$ & \phs$0.84$$\pm$$0.10$ & \phs$2.24$$\pm$$0.11$ & \phs$1.52$$\pm$$0.12$ & $0.12$ & $0.15$ & 25 \\ %& best-observed \\ 
$5$ & $r$ & $(B-r)$ & $-18.94$$\pm$$0.01$ & \phs$0.81$$\pm$$0.11$ & \phs$0.90$$\pm$$0.08$ & \phs$1.46$$\pm$$0.11$ & $0.12$ & $0.15$ & 25 \\ %& best-observed \\ 
$6$ & $i$ & $(B-V)$ & $-18.49$$\pm$$0.02$ & \phs$0.56$$\pm$$0.11$ & \phs$1.05$$\pm$$0.11$ & \phs$1.68$$\pm$$0.12$ & $0.12$ & $0.15$ & 25 \\ %& best-observed \\ 
$7$ & $Y$ & $(B-V)$ & $-18.45$$\pm$$0.01$ & \phs$0.39$$\pm$$0.12$ & \phs$0.55$$\pm$$0.11$ & \phs$1.80$$\pm$$0.17$ & $0.10$ & $0.15$ & 19 \\ %& best-observed \\ 
$8$ & $J$ & $(V-J)$ & $-18.34$$\pm$$0.03$ & \phs$0.57$$\pm$$0.08$ & \phs$0.17$$\pm$$0.06$ & \phs$1.45$$\pm$$0.25$ & $0.04$ & $0.13$ & 20 \\ %& best-observed \\ 
$9$ & $H$ & $(V-H)$ & $-18.21$$\pm$$0.04$ & \phs$0.28$$\pm$$0.15$ & \phs$0.20$$\pm$$0.08$ & \phs$2.75$$\pm$$1.74$ & $0.12$ & $0.15$ & 18 \\ %& best-observed \\ 
$10$ & $K$ & $(V-K)$ & $-18.34$$\pm$$0.04$ & \phs$0.79$$\pm$$0.29$ & \phs$0.15$$\pm$$0.10$ & \phs$4.15$$\pm$$8.48$ & $0.10$ & $0.17$ & 9 \\ %& best-observed \\ 
\enddata 
\tablecomments{Fits of the type: $\mu_X=m_X^{\mathrm{max}}-M_X(1.1,0)-b_X\,[\dm-1.1]-\beta_X^{YZ}\, (Y-Z).$ \\ Columns: (1) Fit identifier; (2) Filter corresponding to $m_X^{\mathrm{max}}$; (3) Color; (4) Absolute magnitude for $\dm=1.1$ and zero $(Y-Z)$ color; (5) Luminosity-decline rate slope; (6) Luminosity-color slope; (7) Corresponding parameter $R_V$ of the CCM+O reddening law; (8) Resulting intrinsic dispersion of SN data; (9) RMS of fit in magnitudes; (10) Number of SNe used in fit.} %; (11) Sample of SNe used in fit (see text).}

\end{deluxetable} 

\label{tab8}
\clearpage
\begin{deluxetable}{lcccc}
\tabletypesize{\normalsize}
\tablewidth{0pt}
\tablecolumns{4}
\tablenum{9}
\tablecaption{Near-IR absolute magnitudes at maximum\tablenotemark{a}\label{tab9}}
\tablehead{
\colhead{Group} &
\colhead{Filter} &
\colhead{$\langle$M$\rangle$} &
%\colhead{$\sigma _x$} &
\colhead{N} 
}
\startdata
Peak early: &              &                         &           \\
            &    $J$       &  $-$18.611 $\pm$ 0.033  &       23  \\
            &    $H$       &  $-$18.318 $\pm$ 0.029  &       23  \\
            &    $K$       &  $-$18.442 $\pm$ 0.033  &       21  \\
            &              &               &            \\
Peak late:  &              &               &            \\
            &    $J$       &  $-$17.847 $\pm$ 0.050  &      4  \\
            &    $H$       &  $-$17.895 $\pm$ 0.013  &      4  \\
\enddata
\tablenotetext{a}{
Normal SNe~Ia typically peak in the near-IR 3 days before
T($B$)$_{\rm max}$.  SNe~Ia that peak in the near-IR a few days after T($B$)$_{\rm max}$
are fast decliners at optical wavelengths and are faint in all bands.}
\end{deluxetable}

\label{tab9}
\clearpage
\begin{deluxetable} {cccc}
\tabletypesize{\normalsize}
\tablewidth{0pt}
\tablecolumns{4}
\tablenum{10}
\tablecaption{Equivalent Widths and Velocities of Na~I~D Lines\label{tab10}}
\tablehead{
\colhead{} &
\colhead{Heliocentric} &
\multicolumn{2}{c}{Equivalent Width (\AA)} \\
\colhead{System} &
\colhead{Velocity (km~s$^{-1}$)} &
\colhead{D1} &
\colhead{D2}} 
\startdata
\multicolumn{4}{c}{\bf 2006dd}\\
A & 1380~$\pm$~15 & 0.62~$\pm$~0.04 & 0.73~$\pm$~0.03 \\
B & 1587~$\pm$~15 & 0.89~$\pm$~0.04 & 0.98~$\pm$~0.05 \\
\hline
\multicolumn{4}{c}{\bf 2006mr}\\
A & 1466~$\pm$~12 & 0.44~$\pm$~0.03 & 0.51~$\pm$~0.02 \\
B & 1585~$\pm$~12 & 0.81~$\pm$~0.03 & 0.62~$\pm$~0.03 \\
\enddata
\end{deluxetable}

\label{tab10}

\clearpage

\clearpage

\clearpage

\clearpage

\clearpage

\clearpage

\clearpage

\clearpage

\clearpage

\clearpage

\clearpage

\clearpage

\clearpage

\end{document}